\documentstyle{mn}

\newcommand{\B}{{{B}}}
\newcommand{\g}{_{\rm g}}
\newcommand{\HH}{H}
\newcommand{\Ko}{{\cal K}}
\newcommand{\nel}{n_{\rm e}}
\newcommand{\phminus}{\phantom{-}}

\newcommand{\RM}{{\rm RM}}

\newcommand{\cmcube}{\,{\rm cm^{-3}}}

\newcommand{\kpc}{\,{\rm kpc}}

\newcommand{\mkG}{\,\mu{\rm G}}

\newcommand{\p}{\,{\rm pc}}
\newcommand{\radm}{\,{\rm rad\,m^{-2}}}

\newcommand{\solar}{_{\odot}}
\newcommand{\half}{{\textstyle{1\over2}}}

\input epsf
\epsfverbosetrue

\title{Structures in the RM sky}
\author[P.~Frick et al.]
       {P.~Frick,$^1$ R.~Stepanov,$^1$ A.~Shukurov$^{2}$ and
D.~Sokoloff$\,^{3}$\\
$^1$Institute of Continuous Media Mechanics,
Korolyov str.~1, 614061 Perm, Russia \\
$^2$Department of Mathematics, University of Newcastle, Newcastle upon
Tyne NE1~7RU, U.K.\\
$^3$Department of Physics, Moscow University, 119899, Moscow, Russia
}
\date{Accepted ....
      Received ...;
      in original form ...}

\begin{document}

\maketitle
\label{firstpage}
\pagerange{\pageref{firstpage}--\pageref{lastpage}}

\begin{abstract}
Coherent structures in the distribution of the Faraday rotation
measure of extragalactic radio sources are isolated using wavelet
transformation technique. A new algorithm of wavelet analysis for
data points nonuniformly distributed on a sphere is developed and
implemented. Signatures of the magnetic fields in the local (Orion)
arm, the Sagittarius arm (and its extension, the
Carina arm), the synchrotron Loop~I and, possibly, the Perseus arm
have been revealed using the $\RM$ catalogues of Simard-Normandin et
al.\ (1981, 551 source) and Broten et al.\ (1988, 663 sources).
Unlike earlier analyses of the $\RM$ sky, our approach has allowed us
to assess the stability of the results with respect to modifications
of the data sample. Only the aforementioned features remain stable
under mild sample modifications.

We consider separately low-latitude sources at $|b|<10^\circ$ and,
using the model of electron density distribution of Cordes et al.\
(1991), we estimate magnetic field strength by comparing the model
wavelet transform with that of the real data.  Independent estimates
of the mean magnetic field strength in the Orion arm using low- and
high-latitude sources converge to $1.4\pm0.3\mkG$.  Rotation measures
of low-latitude sources provide a clear indication of a magnetic
field reversal at a distance 0.6--1\,kpc towards the Galactic centre.
Our analysis has revealed for the first time the extension of the
reversal in the Carina arm.  Low-latitude sources from
the catalogue of Broten et al.\ (1988)  indicate a magneto-ionic
structure in the direction of the Perseus arm with the magnetic field
direction reversed with respect to that in the Orion arm.
The extent of the region with reversed magnetic field near the
Sun is 3\,kpc or more in the azimuthal direction. The average pitch
angle of magnetic field in the nearby spiral arms is $15^\circ$, and
the mean field strength in the Sagittarius--Carina and Perseus arms
is $1.7\pm0.3\mkG$ and $1.4\pm1.2\mkG$, respectively.  The
line-of-sight magnetic field in Loop~I is estimated as
$0.9\pm0.3\mkG$.  

We find firm evidence of a dominant even symmetry of the local mean
magnetic field with respect to the galactic equator.  Our results are
compatible with a moderate large-scale north-south asymmetry, with
magnetic field in the southern hemisphere being stronger in a region
of at least $3\kpc$ in size.  It cannot be excluded, however, that the
asymmetry is local and results from vertical bending of magnetic
lines in a region of about $400\p$ in size, with the Sun being
located close to the top of a magnetic loop whose magnetic field is
$0.5\mkG$ stronger than the average field.
\end{abstract}


\section{ Introduction}                 \label{Intro}

Faraday rotation measures of polarized extragalactic radio sources
and pulsars are a direct and informative tracer of the large-scale
magnetic field of the Milky Way (see the reviews of Spoelstra 1977,
Zeldovich, Ruzmaikin \& Sokoloff 1983, Vall\'ee 1983a, Sofue,
Fujimoto \& Wielebinski 1986, Ruzmaikin, Shukurov \& Sokoloff 1988,
Wielebinski \& Krause 1993, Kronberg 1994, Beck et al.\ 1996, Zweibel
\& Heiles 1997, Vall\'ee 1997).  Faraday rotation measure $\RM$ is a
weighted integral of the longitudinal magnetic field along the line
of sight,
\begin{equation}
\RM=K\int_0^L \nel {\bmath{\HH}}\cdot d{\bmath{s}}\;,
\label{RM}
\end{equation}
where $\nel$ is the number density of thermal electrons,
$\bmath{\HH}$ is the magnetic field, $L$ is the distance to the radio
source and $K=0.81\,{\rm rad\,m^{-2}\,cm^3\,\mu G^{-1}\,pc^{-1}}$
(the positive direction of $\bmath{s}$ is that from the source to the
observer, so that field directed towards the observer produces
positive $\RM$).  Since thermal electron density is relatively
small in the intergalactic space, the main contribution to (\ref{RM})
for extragalactic sources comes from the Milky Way and from the radio
source itself. As magnetic fields in different radio sources are
uncorrelated, the distribution of $\RM$ in the sky for a
representative sample of radio sources can be used to study the
magnetic field in our Galaxy.

For extragalactic radio sources, the effective value of $L$ is
determined by the angular coordinates of the source in the sky,
whereas Faraday rotation measures of pulsars essentially depend on
the pulsar position within the Galaxy. Therefore, Faraday rotation
measures of pulsars and extragalactic radio sources are most often
considered separately, although one can envisage procedures of their
joint analysis.  In this paper we discuss the extragalactic $\RM$ sky
and those features of the Galactic magnetic field that can be deduced
from its analysis.  We apply wavelet transform techniques, a method
perfectly suited to isolate regular structures in a noisy signal.

\section{Studies of the RM sky}        \label{SRMS}

Early analyses of the configuration of the large-scale Galactic
magnetic field using Faraday rotation measures of extragalactic radio
sources were performed by Morris \& Berge (1964) (37 sources, not all
of them extragalactic) and Gardner, Morris \& Whiteoak (1969) (355
sources). These authors noted a strong asymmetry between the northern
and southern Galactic hemispheres and concluded that the average
magnetic field in the Solar neighbourhood is directed toward the
Galactic longitude $l=70$--$80^\circ$. Morris \& Berge (1964) claimed
that the field has opposite directions above and below the Galactic
midplane.  Gardner et al.\ (1969) suggested that the apparent
antisymmetry is due to a local anomaly at $|l|<60^\circ$ associated
with the radio Loop~I. These authors also noted that the apparent
direction of the field shifts to larger $l$ with the increase of the
latitude of the sources considered; we explain this behaviour below
in Sect.~\ref{LLS}.  Vall\'ee \& Kronberg (1973) measured rotation
measures of 252 sources (published by Vall\'ee \& Kronberg 1975) to
estimate the direction of the field as $l=90^\circ\pm10^\circ,\
b=5^\circ\pm10^\circ$ (with the region of Loop~I excluded from the
analysis) and concluded that the large-scale field is similarly
directed above and below the Galactic plane.  Ruzmaikin \& Sokoloff
(1977) considered 154 sources from the sample of Morris \& Tabara
(1973) and estimated the field direction as about $l=99^\circ$ (with
the Loop~I region excluded) presuming that the field direction is the
same above and below the midplane.  Ruzmaikin, Sokoloff \& Kovalenko
(1978) considered the catalogues of Morris \& Tabara (1973) and
Vall\'ee \& Kronberg (1975) (with sources having $|\RM|>100\radm$
removed); these authors have determined the field
direction in the southern Galactic hemisphere to be
$l=91^\circ\pm22^\circ$ and stressed that it is not possible to
determine reliably the global field direction in the northern
hemisphere because of the local distortions due mainly to
Loop~I.

Comprehensive catalogues of the Faraday rotation measures of
extragalactic radio sources were published by Tabara \& Inoue (1980),
Eichendorf \& Reinhardt (1980), Simard-Normandin, Kronberg \&
Button (1981) and Broten, MacLeod \& Vall\'ee (1988), which contain
770, 346, 555 and 674 sources, respectively.  $\RM$ data were also
obtained for several localized regions in the sky (see Vall\'ee
1983a). The data used in this paper are discussed in detail in
Sect.~\ref{DDS}.

Simard-Normandin \& Kronberg (1979, 1980) used RMs for samples of 476
and 543 extragalactic radio sources, respectively. These authors
averaged RMs of neighbouring sources within a $10^\circ$--$15^\circ$
radius of each source in order to reduce the noise. They also noted
that the RM distribution in the southern Galactic hemisphere is
significantly more ordered than that in the northern hemisphere. The
largest localized feature is Loop~I which produces a perturbation of
about $30\radm$ at $b>+20^\circ$ in the first and fourth Galactic
quadrants.  The  pitch angle of the local large-scale magnetic field
is about $15^\circ$, as estimated from the fact that the associated
positive RMs of the sources at $|b|>5^\circ$ have a peak at about
$l=255^\circ$ (instead of $270^\circ$) and $\RM$ changes sign at
about $l=165^\circ$. These authors argued that a region of positive
$\RM=50$--$300\radm$ centred at $l\simeq40^\circ,\ b\simeq+5^\circ$
is due to a reversed magnetic field in the Sagittarius arm. These
authors were unable to identify the reversal in the extension of the
Sagittarius arm to the fourth Galactic quadrant (the Carina arm)
using their methods.

Andreassian (1980) considered 301 extragalactic sources with 251 of
them taken from Vall\'ee \& Kronberg (1975) and the remaining 50 from
Haves (1975) and excluded those having $|\RM|>100\radm$ at
$15^\circ<|b|\leq75^\circ$ and $|\RM|\leq8\radm$, so that only 176
sources remained. He considered least-squares fits using a model for
the local large-scale magnetic field to conclude that the field has
opposite directions above and below the Galactic plane. Andreassian
(1982) used 419 sources from the catalogue of Tabara \& Inoue (1980)
applying the same selection criteria (retaining the region of Loop~I)
to conclude that the fields revealed by the sources at $b>15^\circ$
and $b<15^\circ$ have opposite directions.

Using the $\RM$ data published by Tabara \& Inoue (1980), Inoue \&
Tabara (1981) estimated the field direction as $l=100\pm10^\circ$
from 224 sources located at $|b|>30^\circ$ and $l\simeq80^\circ$ from
those at $|b|<30^\circ$.
Sources with $|\RM|>100\radm$ were excluded from
their analysis. The original rotation measures were averaged over
circles of a $15^\circ$ radius centred on each source.  These authors
did not confirm the presence of a reversal of the regular magnetic
field between the Orion and Sagittarius spiral arms\footnote{The
feature called here the Orion arm is plausibly a relatively short
spur between the more extended Sagittarius and Perseus arms, so that
many authors call it the Orion spur.} and noted that the apparent
reversal suggested by Simard-Normandin \& Kronberg (1980) may be due
to Loop~I.

Sofue \& Fujimoto (1983) also analyzed the sample of Tabara and Inoue
(1980) consisting of 1510 extragalactic radio sources 770 of which
have known $\RM$, but applied less stringent selection criteria. They
discarded an unspecified number of sources with $|\RM|>300\radm$ and
with errors exceeding $10\radm$. The authors then used
Gaussian-weighted mean values of $\RM$ for each source and studied
features outside the local arm taking the  averaging Gaussian with a
full width at half-maximum of $20^\circ$. They
found four alternating maxima and minima of
$\RM$ along Galactic longitude at $|b|<20^\circ$
and noted that they approximately
coincide with tangential directions to the inner Galactic spiral
arms, and interpreted this as an evidence of the corresponding number
of magnetic field reversals in the inner Galaxy.

Attempts to determine the field direction in the next outer Galactic
spiral arm, the Perseus arm, were made by Vall\'ee (1983c) and
Agafonov, Ruzmaikin \& Sokoloff (1988) who used 45 sources located in
this region of the sky. Although both papers used the same RM data,
they came to opposite conclusions about the direction of the field in
the Perseus arm: Vall\'ee argues that it is parallel to that in the
Orion arm, whereas Agafonov et al.\ conclude that the fields are
almost antiparallel.

Clegg et al.\ (1992) obtained and analyzed Faraday rotation measures
of 56 low-latitude extragalactic sources at $45^\circ<l<93^\circ,\
|b|<5^\circ$ and compared their longitudinal distribution with
predictions of various magnetic field models. They adjusted
parameters of a circular-field model to argue that the field has a
reversal between the Sagittarius and Orion arms producing a null in
$\RM$ at $l=62^\circ$ with $\RM$ mainly positive at $l<62^\circ$ and
negative at $l>62^\circ$.

Han \& Qiao (1994) used Faraday rotation measures of 103
extragalactic sources at $|b|>60^\circ$ from the compilation of
Broten et al.\ (1988) in an attempt to estimate the
vertical magnetic field (sources with $|\RM|\geq300\radm$ were
excluded from their analysis). Their estimate is
$B_z\simeq0.2$--$0.3\mkG$ with a direction from the south to the
north Galactic pole. It is, however, unclear how one can reliably
distinguish the local north-south asymmetry associated, e.g., with
Loop~I from a global vertical magnetic field.

A recent study of the regular magnetic field of the Galaxy using
rotation measures of extragalactic radio sources was performed by Han
et al.\ (1997). These authors used the catalogue of Broten et al.\
(1988), omitted sources with $|\RM|>250\radm$, averaged the rotation
measures within circles of a $15^\circ$ radius around each source and
discarded those whose $\RM$ deviates strongly from the mean value in
its circle. Altogether, 551 sources were considered.  They noted
that, in addition to the dominance of negative rotation measures at
$0<l<180^\circ$ and positive RMs at $180^\circ<l<360^\circ$ resulting
from the local large-scale magnetic field, the distribution of RM
in the first and fourth quadrants ($-90^\circ<l<0$ and
$270^\circ<l<360^\circ$) is apparently antisymmetric with respect to
the Galactic equator (this is not confirmed by our results); they
found a confirmation of the antisymmetry in the RM distribution for
high-latitude pulsars ($|b|>8^\circ$).  These authors argued that
this pattern (that includes the region of Loop~I) results not from a
local magnetic perturbation but rather from the odd symmetry of the
global magnetic field in the thick Galactic disc within a few
kiloparsecs from its centre.

A general feature of the above analyses is that they provide rather
divergent results concerning magnetic field directions in remote
areas such as neighbouring spiral arms, although the estimates of the
local regular magnetic field are more coherent. Furthermore, most
authors who did not {\it assume\/} the even parity of the magnetic
field with respect to the Galactic equator were unable to find
convincing evidence of this symmetry in the observational data.
These difficulties arise because it is unclear how sensitive the
results are to the selection criteria, to the data sample chosen, and
to the analysis technique applied, and because it is not easy to
isolate the effects of the local perturbations associated mainly with
Loop~I.

The above analyses were based either on fitting a chosen {\it
global\/} model of the large-scale magnetic field to the observed
distribution of Faraday rotation measures or merely on naked-eye
adjustments of the model parameters.  This approach can be improved
in the following respects.  The data themselves are mostly sensitive
to the magnetic field near the Sun because extragalactic sources with
known $\RM$ are observed mainly at high galactic latitudes. Even
nearby spiral arms (Sagittarius and Perseus)
occupy in the sky regions as small as $|b|\leq5$--$10^\circ$ in the
directions tangent to the arms (assuming the scale height of 500\,pc
for the magneto-ionic layer).  This difficulty cannot be completely
removed by improving statistics and enriching the sample with
low-latitude sources since it is difficult to distinguish remote
large-scale features from nearby localized magnetic perturbations.
(Pulsar data are more favourable in this respect provided distances
to the pulsars are reliable.) Therefore, one can expect significant
difficulties in deciding between models of the magnetic field having
similar properties near the Sun but differing in global features
(e.g., field reversals along the galactic radius and axisymmetric
versus bisymmetric geometry).  Clearly, in this situation it is
important to use well-substantiated, model-independent methods of
quantitative statistical analysis rather than simple naked-eye fits.

A natural way to avoid these difficulties is to study, as a first
step, the magnetic field in a close vicinity of the Sun as this field
is described by an irreducibly small number of independent
parameters.  Ruzmaikin et al.\ (1978) applied a regression method to
obtain estimates of the strength and direction of the magnetic field
near the Sun from Faraday rotation measures of extragalactic radio
sources.  However, they used first-generation RM catalogues which
contain a rather small number of sources and many unreliable
measurements. Ryadchenko \& Sokoloff (1997, unpublished) applied this
method to the more recent catalogue of Simard-Normandin et al.\
(1981) to obtain basically the same results as Ruzmaikin et al.\
(1978).  Appropriate statistical analyses were performed for pulsar
data by Rand \& Kulkarni (1989) and Rand \& Lyne (1994) who fitted
the data with models of local and global large-scale magnetic fields.

To study the global structure of the Galactic magnetic field in an
objective manner without prescribing any model of magnetic field
beforehand, one may use methods similar to Fourier decomposition,
that is expanding the observed spatial distribution of $\RM$ over a
complete set of orthogonal basis functions. For spherical
geometry, expansion over spherical harmonics seems to be natural.
Two-dimensional spherical harmonic analysis using 65 extragalactic
radio sources was performed by Seymour (1966, 1984) who concluded
that the harmonics of the first four orders are statistically
significant, and that the dominant toroidal component of the Galactic
magnetic field is symmetric with respect to the Galactic equator; the
local direction of the magnetic field was determined to be from
$l=112^\circ$ to $l=292^\circ$ (probably a misprint: the opposite
direction was meant by the author).  Seymour also concluded that
there is a weaker antisymmetric toroidal field increasing with
distance from the Galactic equator, which the author identified with a
halo field.

Spherical harmonic analysis is not an optimal tool for
studies of the global magnetic field of the Galaxy, again because
more or less distant parts of the Milky Way occupy small regions in
the sky and would require the inclusion of high-order spherical
harmonics into the fit.  A generic problem with Fourier expansion is
that the basis functions (e.g., spherical harmonics) are spread over
the whole sphere (i.e., they are not localized in space).  Therefore,
small-scale features are represented by a large number of phased
Fourier modes.  Statistical noise in the data and uneven distribution
of the data points in the sky hamper reliable estimates of the
required large number of the fit parameters.

In this paper we apply wavelet analysis to the distribution of
Faraday rotation measures of extragalactic radio sources in the sky.
Our aim is to understand which particular parameters of the Galactic
magnetic field can be reliably extracted from the available data and
then to estimate these parameters.  Unlike previous analyses, we also
verify the stability of our results with respect to modest
modifications of the sample in order to isolate those features in the
$\RM$ sky which are less sensitive to the incomplete sampling. Our
main goal here is not to present an interpretation of the observed
$\RM$ distribution in terms of competing theories of Galactic
magnetic fields but rather to clarify which details in the $\RM$ sky
are real and which can be artifacts of irregular data grids,
incomplete sampling, diverse selection criteria, etc.  Results of the
wavelet analysis of the $\RM$ sky and their interpretation in terms
of interstellar magnetic fields are discussed in Sect.~\ref{SIFRS}
and \ref{MFOA}.

Wavelet analysis (see, e.g., Farge 1992, Holschneider 1995a) is a
natural generalization of the Fourier analysis aimed at a consistent
description of not only the scale but also of the spatial
localization of structures. Unlike decompositions in harmonic
functions, wavelet analysis uses basis functions that are
localized in both physical and wave number spaces. This allows to
represent localized structures in the data in terms of a tractably
small number of parameters.  Wavelet analysis is therefore a perfect
tool for objective studies of the RM sky. We discuss our
implementation of wavelet analysis and in Sect.~\ref{WTA} and
Appendix~A.

Wavelet analysis has been introduced in its present form by Grossmann
\& Morlet (1984) for analysis of seismic time series. Its physical
applications are abundant, including theory of turbulence (Farge
1992), where this method was independently introduced by Zimin (1981)
(see also Frick \& Zimin 1993), and now it is also used in
astrophysics, e.g. for studies of large-scale structures in the
Universe (Bijaoui, Slezak \& Mars 1993; Malik \& Subramanian 1997),
cosmological velocity fields (Rauzy, Lachi\`eze-Rey \& Henriksen
1995), solar (Nesme-Ribes et al.\ 1995; Lawrence, Cadavid \&
Ruzmaikin 1995; Halser et al.\ 1997; Frick et al.\ 1997b) and stellar
activity (Soon, Frick \& Baliunas 1999), and galactic magnetic fields
(Frick et al.\ 2000).

\section{Data selection}   \label{DDS}

The catalogue of Simard-Normandin et al.\ (1981) contains carefully
selected values of $\RM$ for 555 unresolved extragalactic radio
sources.  This catalogue is rather extensive and is notable for
well-defined and consistent criteria of data selection.
Most sources in the catalogue of Simard-Normandin et al.\ have
$|\RM|<200$--$300\radm$. It is generally believed that large values
of $|\RM|$ of the order of, say, $500\radm$ in such catalogues (based
on observations with only moderate wavelength coverage) are hardly
due to the contribution of the interstellar magnetic field but rather
arise from magnetic fields in the sources, unreliable determinations
of $\RM$ due to the $n\pi$ ambiguity in polarization angles,
nonlinear relation between the polarization angle and the wavelength
squared, etc.\ (Ruzmaikin \& Sokoloff 1979, Vall\'ee 1980).

The catalogue of Broten et al.\ (1988) (we used its
update of 1991 kindly provided to us by Dr.\ J.~P.~Vall\'ee) is based
on similar methods and selection criteria; it contains data for 674
extragalactic radio sources. The catalogue contains a number of
sources with very large $|\RM|$, and many of them are at low Galactic
latitudes. Together with the catalogue of Simard-Normandin et al.\
(1981), this catalogue, especially its low-latitude sources, provides
the basis of our analysis.

The catalogues of Tabara \& Inoue (1980) and Eichendorf \& Reinhardt
(1980) contain a number of radio sources which are not present in the
catalogues of Simard-Normandin et al.\ (1981) and Broten et al.\
(1988). Further results have been published for radio sources in
selected regions in the sky:  the Perseus arm (Vall\'ee 1983b),
the Scutum arm (Vall\'ee, Simard-Normandin \& Bignell 1988), the
Monogem arc (Vall\'ee, Broten \& MacLeod 1984) and the Gum nebula
(Vall\'ee \& Bignell 1983), and also for other sources at low Galactic
latitudes (Clegg et al.\ 1992).
Much of these data have been included into the catalogue of
Broten et al.\ (1988).  It is not a trivial matter to combine all
the data into a larger joint catalogue because different authors use
different selection criteria and procedures for the determination of
$\RM$ from observations.  In particular, the catalogues of Tabara \&
Inoue and Eichendorf \& Reinhardt contain a significantly larger
fraction of sources with very large values of $|\RM|$ than the
catalogues of Simard-Normandin et al.\ and Broten et al.

There are some contradictions between these catalogues, so we have
selected the data in the following manner. If the other catalogues
disagree with that of Simard-Normandin et al.\ or Broten et al., we
usually preferred the data from the latter two catalogues. Some
exceptions were made when  these two catalogues give a significantly
larger value $|\RM|$ for a source than the other catalogues.  We
discarded such sources.

The list of Clegg et al.\ contains many resolved sources with very
large $|\RM|$, and the authors could not find a unique value of RM
for some of them. We neglected these data as well.  After such a
selection, only seven sources remain in the list of Clegg et al. As
it seems to be more important to preserve the uniformity of the data
set than to extend it insignificantly, we excluded also these seven
estimates from our analysis.

We also removed the sources which are not mentioned by
Simard-Normandin et al.\ or Broten et al.\ and for which the other
catalogues give different estimates of $|\RM|$ all exceeding
$100\radm$. However, if one of the catalogues gives an estimate lower
than $100\radm$, we accepted it.

The above procedure yields a joint catalogue of Faraday rotation
measures. The catalogue can be found at the URL
http://plaxa.icmm.ru/\~{}lab4/RMjoint.dat.

The total number of sources in the joint catalogue is 841. Only 4
sources have been discarded from the catalogue of Simard-Normandin et
al.\ and 41 source omitted from the catalogue of Broten et al.
However, we appreciate that the joint catalogue still may contain
inconsistent data. We also stress again the importance of the
uniformity of the data set to be used.

Our results are based on the catalogues of Simard-Normandin et al.\
(1981) (551 sources have been used) and Broten et al.\ (1988; 663
sources). Magnetic field parameters for the remote parts of the
Galaxy have been derived from low-latitude sources represented mainly
in the latter catalogue.  We compared the results obtained from the
catalogues of Simard-Normandin et al., Broten et al.\ and the joint
catalogue to check the  stability of our conclusions with respect to
variations in the data set.  We have also checked the stability of
the results by removing sources with large $|\RM|$ from these
catalogues.

We consider only those magneto-ionic structures whose angular radius
noticeably exceeds $20^\circ$ (at least along Galactic longitude),
the resolution limit of our technique (see Appendix~A). Analysis of
structures at smaller angular scales using our approach would require
a catalogue with a significantly more uniform and dense sky coverage.
Therefore, we restrict our attention to the nearest spiral arms and
the largest single radio structure, Loop~I. These features are stable
with respect to modest modifications of the data. Structures of a
smaller angular size (more distant spiral arms and magnetic bubbles)
can be revealed by specialized detailed analysis of selected zones
(e.g., Vall\'ee et al.\ 1988; Vall\'ee 1983a, 1993, 1997).

\section{Wavelets: a scale filtering technique}   \label{WTA}

Wavelet analysis is based on a convolution of the data with a family
of basis functions that depend on two parameters, the scale and
location. This is a generalization of the Fourier transformation.  In
the latter, harmonic functions are used as a functional basis
characterized by a single parameter, the frequency.  In both wavelet
and Fourier transforms, the basis functions have zero mean values.
However, wavelet transformations use oscillatory functions
decaying at infinity, i.e.\ localized in space. This introduces an
additional parameter, the position vector of the centre of a wavelet.
In order to avoid excessive number of parameters, a family of
wavelets is constructed from self-similar functions. The
self-similarity can be implemented exactly in flat space (e.g., on a
plane), but only approximately in curved space, e.g.\ on a sphere
(Torresani 1995; Holschneider 1995b).

The self-similarity distinguishes wavelet transformation from the
windowed Fourier transformation, where the frequency, the width of
the window and its position are independent parameters. These basis
functions are not self-similar and, in an isotropic case, contain
three parameters (one of them, the window position, can be a vector),
which complicates both interpretation of the results and the
construction of the inverse transformation.  Therefore, windowed
Fourier transforms are usually applied only in the simplest
one-dimensional case (e.g., to time series).

We consider continuous wavelet transformation in two dimensions on a
sphere of unit radius (the sky). We introduce angular coordinates $l$
and $b$, the galactic longitude and latitude, respectively.  The
wavelet transform $w(a,l,b)$ of the signal $\RM(l,b)$ is defined as
\begin{equation}
w(a,l,b) =\int\!\!\!\int
\psi_{a,{\bmath{r}}}({\bmath{r}}')
\RM({\bmath{r}}')\,d^2{\bmath{r}}'\;,
\label{wt}
\end{equation}
where $\psi_{a,{\bmath{r}}}$ is the wavelet of a scale $a$ centred at
a position ${\bmath{r}}=(l,b)$ on the sphere (${\bmath{r}}$ should
not be confused with the radius vector) and the integration is over
the unit sphere.  The wavelet transform can be understood as the
amplitude (positive or negative) of a structure of a scale $a$ at a
position $(l,b)$. Wavelets must have zero mean value,
\begin{equation}
\int\!\!\!\int\psi_{a,{\bmath{r}}}({\bmath{r}}')\,d^2{\bmath{r}}' =
0\;.
\label{w-adm}
\end{equation}

The choice of the wavelet depends on the goals of the analysis. For a
local spectral analysis in time or space, wavelets with a good
spectral resolution (i.e., well localized in Fourier space, or having
many oscillations) should be used.  To localize and analyze spatial
structures represented in the data, which is our goal here, a wavelet
which is well localized in the physical space is preferable (i.e.,
rapidly decaying at infinity).

\epsfxsize=6cm
\begin{figure}
\centerline{\epsfbox{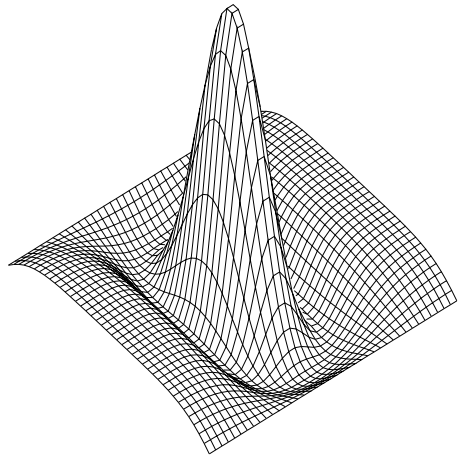}}
\caption[]{The `Mexican hat' wavelet in the $(l,b)$-plane
($c=2$).}
\label{hat_fig}
\end{figure}

We use a simple real isotropic wavelet known as the {\it Mexican hat\/}
\begin{equation}
\psi(x) = \left(c-x^2\right)\exp{\left(-\half x^2\right)}\;,
\label{hat}
\end{equation}
where $c$ is chosen to ensure that the mean value of $\psi(x)$ is
zero; $c=1$ in one dimension; the `Mexican hat' in the plane, shown
in Fig.~\ref{hat_fig}, has $c=2$.  The wavelet family is generated
from a function $\psi(x)$, called the analyzing wavelet, by its
dilation and translation characterized by two parameters, the scale
$a$ and position $\bmath{r}$ of a wavelet. A family of
two-dimensional isotropic wavelets has the form
\begin{equation}
\psi_{a,{\bmath{r}}}({\bmath{r}}') =
a^{-1} \psi\! \left(s ({\bmath{r}},{\bmath{r}}')/a\right)\;,
\label{w-aaa}
\end{equation}
where ${\bmath{r}}'=(l',b')$ is the current angular position in the
sky and $s$ is the angular distance between two points on unit sphere,
${\bmath{r}}=(l,b)$ and ${\bmath{r}}'=(l',b')$, given by
\begin{equation}
s({\bmath{r,r}}') =
\arccos[\cos{b}\cos{b'}\cos{(l-l')}+\sin{b}\sin{b'}]\;.
\label{dist}
\end{equation}
We generalize the definition of a wavelet family by introducing the
dependence of $c$ on $a$ in Eq.~(\ref{hat}) in order to implement an
approximate self-similarity on a sphere (see Appendix~A for details).
Geometric factors in Eq.~(\ref{dist}) make the wavelet
anisotropic in terms of the angle variables $(l,b)$.

The traditional wavelet transformation should be further modified in
our case because the data are given at irregularly distributed points
in the sky, being neither a continuous function nor given on a
regular grid.  Our implementation, based on the {\it gapped wavelet
technique\/} introduced by Frick et al.\ (1997a), is described in
Appendix~A. This modification also affects $c$ in Eq.~(\ref{hat}):
now $c$ becomes a function of the position of the wavelet centre,
$\bmath{r}$.

We present some of our results in the form of maps of the wavelet
transform $w(a,l,b)$ for a given scale $a$. The scale of a structure
is characterized by the angular distance $R$ from the centre of a
wavelet to its first zero, i.e.\ $\psi|_{x=R}=0$ and
$R(a)=c^{1/2}a$ -- see Eq.~(\ref{hat}).  $R$ as a function of $a$ is
derived and discussed in Appendix~A.  The wavelet transformation was
performed for a wide range of $R$, but for
practical reasons we varied $R$ by steps of $3^\circ$ (as determined
by the typical separation of the data points); thus all the values of
$R$ quoted below have the accuracy of $\pm3^\circ$. We display only
those wavelet transform maps which correspond to physically
distinguished values of $R$. The maps shown below are in fact
scale-filtered $\RM$ distributions with a filtering window centred at
a certain scale $R$.

\epsfxsize=6.3cm
\begin{figure}
\centerline{\epsfbox{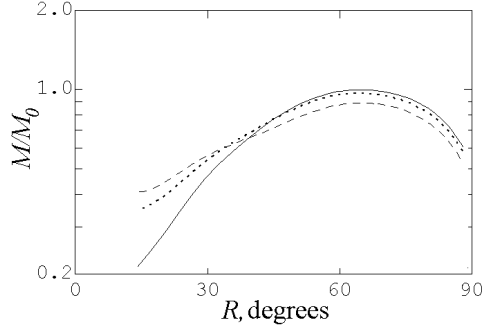}} \caption[]{The integral wavelet
spectra $M(R)$ for the catalogues of Simard-Normandin et al.\
(1981) (solid), Broten et al.\ (1988) (dashed) and the joint
one (dotted). The spectra are normalized by the maximum value of
$M(R)$ for the catalogue of Simard-Normandin et al.}
\label{spectrum}
\end{figure}

Our aim is to isolate spatial structures in the $\RM$ distribution in
the sky and to estimate their parameters. For this purpose we
consider the quantity
\begin{equation}
M(R)={1\over{4\pi}}\int w^2(R,{\bmath{r}})\,d^2{\bmath{r}}\;.
\label{Ma}
\end{equation}
In Fourier analysis, a similar quantity is known as the spectral
energy density. The normalization factor in Eq.~(\ref{w-aaa}) is
chosen so that structures of the same amplitude but different
scales would contribute equally to $M(R)$.  As can be seen in
Fig.~\ref{spectrum}, $M(R)$ has a wide maximum at about
$R=70^\circ$ for all three catalogues used. The maximum in the
spectrum is due to the largest structures in the $\RM$
distribution, that is the magnetic field in the Orion arm (see
Sect.~\ref{SIFRS}). As indicated by a slower decrease of $M$
with $R$, the catalogue of Broten et al.\ and the joint catalogue
contain more structure at small and intermediate scales.

We note that the wavelet used here is devised to obtain a good
spatial resolution at the expense of a poorer resolution in the space
of scales $R$ (or the wave number space).  Therefore, the imprints of
underlying structures may be hardly visible in Fig.~\ref{spectrum},
the more so that $M(R)$ is an integral parameter which smears
individual features even if they are strong.  Such features are
better visible in the wavelet transform maps in the $(l,b)$ plane.

The use of an isotropic wavelet does not preclude the detection of
strongly elongated structures. In such cases the dominant scale $R$
corresponds to the smaller dimension of the structure, and $w$ at
this scale has an elongated maximum. The curvature of an elongated
feature contributes to the wavelet transform at larger scales.

\epsfxsize=8.2cm
\begin{figure}
\centerline{\epsfbox{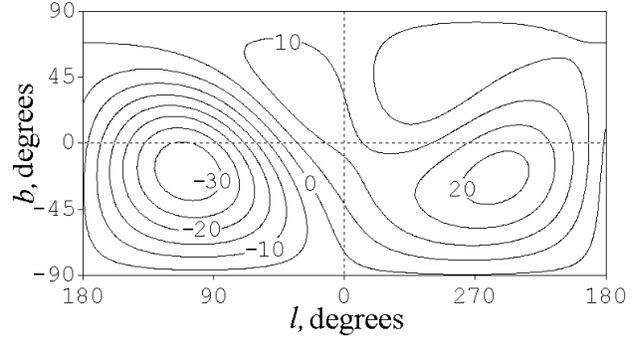}}
\caption[]{Contours of the wavelet transform $w(a,l,b)$ in $\!\radm$
at the dominant scale $R=67^\circ$ for the catalogue of
Simard-Normandin et al.\ (1981). }
\label{main}
\end{figure}

\section{Stable features in the RM sky}\label{SIFRS}
The wavelet transforms presented in this section have been obtained
from the sample of Simard-Normandin et al.\ (1981) unless stated
otherwise. Figure~\ref{main} shows the wavelet transform at the scale
$R=67^\circ$ where $M(R)$ is maximum in all three catalogues (within
$3^\circ$). The main features of this map are two structures located
at about $l=90^\circ$ and $270^\circ$; they have opposite signs.
These maxima are imprints of the large-scale magnetic field in the
local (Orion) arm and Loop~II (the Cetus Arc) (see also Sect.~\ref{MFOA}).
$|\RM|$ has slightly different peak values in the first and fourth
quadrants and the zero-level contour is asymmetric in galactic
longitude. The other two features visible in this map are a mild
maximum of $w$ at $l\simeq30^\circ$ and a minimum at
$l\simeq300^\circ$ in the northern hemisphere. These are
manifestations of Loop~I (see below).

It is notable that the dominant structures at $l\simeq90^\circ$ and $270^\circ$
are shifted from the Galactic equator to negative Galactic latitudes.
The asymmetry between large-scale $\RM$ distributions in the northern
and southern Galactic hemispheres is well known from earlier studies
discussed in Sect.~\ref{SRMS}.

\epsfxsize=8.2cm
\begin{figure}
\centerline{\epsfbox{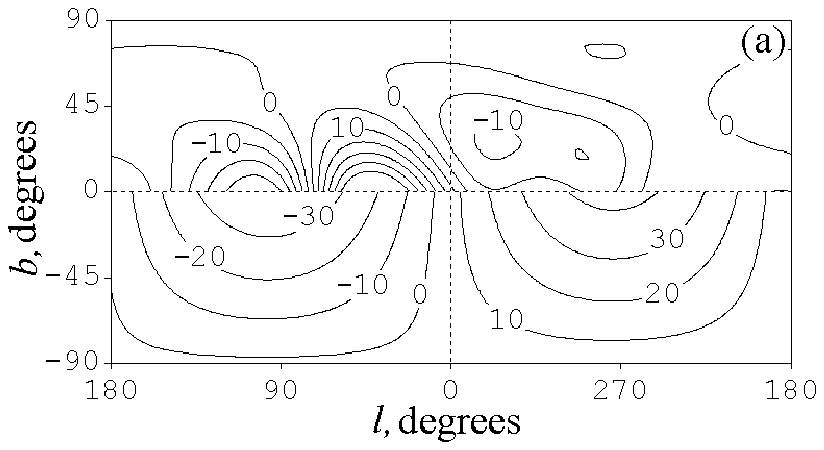}}
\epsfxsize=6.3cm
\centerline{\epsfbox{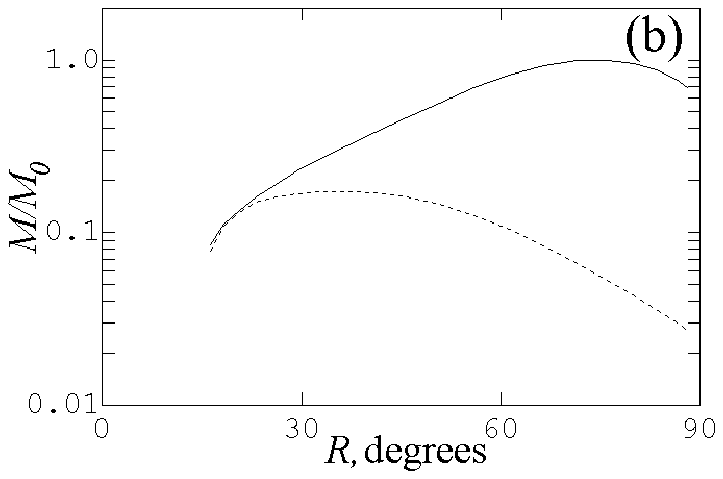}}
\caption[]{Wavelet transforms $w(a,l,b)$ of the sample of
Simard-Normandin et al.\ (1981) obtained separately for the northern
and southern Galactic hemispheres at scales corresponding to the
maxima in their respective energy spectra $M(R)$. (a):  contours of
$w$ (in $\!\radm$) at $R=35^\circ$ for the northern hemisphere (upper
panel) and at $76^\circ$ for the southern hemisphere (lower panel).
(b): $M(R)$ for the southern (solid) and northern (dashed)
hemispheres. $M_0$ is the maximum of $M(R)$ for the southern
hemisphere.  }
\label{hemi}
\end{figure}

In order to quantify the north-south asymmetry we consider for a
moment the two hemispheres separately. In Fig.~4 we show wavelet
transforms and integral spectra for the two hemispheres.  The spectra
are maximum at very different scales, $R=35^\circ$ in the north and
$R=76^\circ$ in the south. This indicates that the magnetic field in
the northern hemisphere is contaminated by distortions whose dominant
angular radius is $35^\circ$. The large-scale features corresponding
to the magnetic field in the Orion arm prevail in the southern
hemisphere, but not in the northern one.

There are two most conspicuous features in the northern hemisphere,
which have no counterparts in the south and which are largely
responsible for the asymmetry. One of them has the centre at
$(l,b)\approx(45^\circ,0)$, and the other is an elongated feature
joining two local maxima at $(l,b)\approx(335^\circ,30^\circ)$ and
$(l,b)\approx(295^\circ,15^\circ)$. These structures are in the
region of Loop~I (and Loop~IV) and the Sagittarius arm.  According to
radio continuum observations at $820\,$MHz, Loop~I occupies
approximately the same region in the sky with its centre at
$(l,b)\approx(329^\circ,18^\circ)$ and the angular diameter of about
$116^\circ$ (Berkhuijsen 1971). The remaining structures in the
northern hemisphere visible in Fig.~4 have signs in agreement with
those in the south, so that they can hardly cause a strong
north-south asymmetry.

\epsfxsize=8.2cm
\begin{figure}
\centerline{\epsfbox{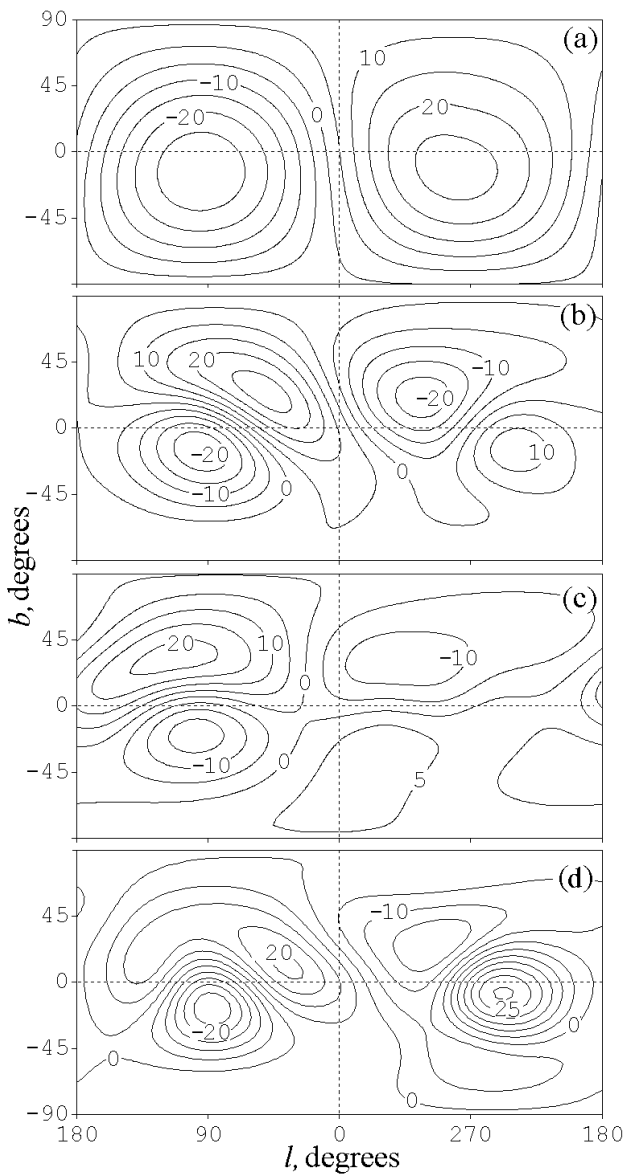}}
\caption[]{Filtered $\RM$ distributions: (a) at a scale $R=76^\circ$
with the region of Loop~I removed; (b), (c) and (d) at a scale
$R=35^\circ$ with the Loop~I region retained but the large-scale
contribution of panel (a) subtracted. The catalogue of
Simard-Normandin et al.\ (1981) was used in (a) and (b), the joint
one in (c) and that of Broten et al.\ (1988) in (d). Contours represent the value of
$w(R,l,b)$ in $\!\radm$.}
\label{Orion}
\end{figure}

In order to isolate the large-scale magnetic field in the Orion arm
we first reduce the north-south asymmetry by omitting all the data in
the region of Loop~I. (Due to an only moderate scale resolution of our
wavelet, the scale filtering, as in Fig.~\ref{main}, is not
sufficient to eliminate contamination by smaller structures.) It is
sufficient to omit points for $l$ from $280^\circ$ to $50^\circ$ at
$b>0$ to match the dominant scales in the two hemispheres. The
wavelet transform was then performed for the whole sphere assuming
that the data in the above region are simply absent.  Our gapped
wavelet technique ensures a correct handling of such spatial
distributions (see Appendix~A).

A wavelet transform map for the dominant scale after removing the
region of Loop~I is shown in Fig.~\ref{Orion}a; it is notable that
the dominant scale remains to be close to $76^\circ$, as for the
southern hemisphere alone.  The map in Fig.~\ref{Orion}a does not
have a gap in the region where we have omitted the data because the
wavelet scale is larger than the size of this region. It is
instructive to compare Fig.~\ref{Orion}a with Fig.~\ref{main} which
shows structures at about the same scale, but for the original data.  The
large-scale magnetic field in the Orion arm is responsible for the
two extrema clearly visible in Fig.~\ref{Orion}a.  The spots are more
symmetric with respect to the Galactic equator than those in the
original data.  We use this map to study the large-scale magnetic
field near the Sun in Sect.~\ref{MFOA}.

In order to study structures at smaller scales, we subtracted from
the original data the component of the scale $R=76^\circ$ discussed
above (i.e., obtained with the data in the region of Loop~I omitted).
Then we wavelet-transformed the result on the whole sphere, including
the region of Loop~I.  This procedure is meaningful provided the
residual obtained after subtracting the dominant component can be
proven to contain a significant signal above the noise level. We
verify this by checking the stability of our results (including those
at $R\simeq30$--$40^\circ$) with respect to sample variations.
Shown in Figs~\ref{Orion}b,c,d are the wavelet transforms at $R=35^\circ$ for the
catalogue of Simard-Normandin et al.\ (1981), the joint one and that of Broten et al.\
(1988), respectively. The distribution of the residual $\RM$ has a wide maximum of
$M(R)$ at $R=35$--$40^\circ$ in all three catalogues.
Apart from small-scale features, the filtered
distributions of the residual $\RM$ at $R=35^\circ$ shown in Figs~\ref{Orion}b,c,d also
reflect the global magnetic field in the Orion arm; the corresponding structures are
located slightly to the south of the Galactic equator. Especially well pronounced and
stable is that at $l\approx90^\circ$.  Its counterpart in the fourth quadrant is weaker
in Figs~\ref{Orion}b,c (this is true also at $R=76^\circ$) and sensitive to the catalogue
chosen: it can be more or less clearly seen at $l\approx250^\circ$ in Figs~\ref{Orion}b,d
but is hardly visible in Fig.~\ref{Orion}c.

The structure of negative $\RM$ seen in Figs~5b,c,d at
$l\approx90^\circ$, $b<0$ occurs in the Region~A of Simard-Normandin
\& Kronberg (1979, 1980) identified with  the Cetus arc (also known
as the synchrotron Loop~II, centred at $l=100^\circ,\ b=-32.5^\circ$
with a diameter of $91^\circ$ -- Berkhuisen et al.\ 1971, Berkhuisen
1971).  The Cetus arc is known to produce a negative contribution to
$\RM$ (Simard-Normandin \& Kronberg 1979, 1980). This makes it
difficult to separate it from the contribution of the Orion arm
that is also negative in that range of $l$.
 We have considered
wavelet transforms at a range of scales $30^\circ<R<67^\circ$, but
the difference in the scale between the two contributions ($67^\circ$
for the Orion arm and presumably about $45^\circ$ for the Cetus arc)
is not sufficient to separate them. Such a separation, based on an analysis of individual
sources, is discussed by Vall\'ee (1982). However, the shift of the wavelet
extremum at $l\approx90^\circ$ to negative latitudes can be, at least in part, due to
the contribution of the Cetus arc. We note in this connection that, at scales about
$R=35^\circ$, the negative extremum at $l\approx90^\circ$ is significantly farther from
the Galactic equator than the positive structure at $l\approx250^\circ$. The $\RM$
structure due to the Cetus arc is plausibly better pronounced at smaller scales, but
global methods similar to that employed here should await new $\RM$ catalogues to isolate
the structure reliably and objectively.

The positive extremum visible in Fig.~\ref{Orion}c at $l\approx 250^\circ$
arises in the region which is close to the Gum nebula centred at $l=260^\circ,\ b=0$
(Vall\'ee \& Bignell 1983, Vall\'ee 1993). The data used here have not allowed us
to reliably identify this feature as its radius, $\simeq20^\circ$ is
close to the resolution of our analysis. This feature is largely due to a few strong
sources in the catalogue of Broten et al.\ located in this area. A
possible signature of the Gum nebula can be seen as an enhancement in the magnitude of
the positive extrememum of the wavelet transform to $35\radm$ at scales $R<40^\circ$ (as
compared to $20\radm$ at larger scales), indicating a narrow intense peak on top of a
wider maximum. The position of this extremum at $l\approx250^\circ,\ b\approx-10^\circ$
changes only slightly as $R$ varies.

There are more details of a scale $R=35^\circ$ located in the
northern hemisphere. Clearly seen in Fig.~\ref{Orion}b are a positive
extremum centred at $(l,b)=(50^\circ,21^\circ)$ and a negative one at
$(l,b)=(289^\circ,23^\circ)$. The position and intensity of the
former structure are different for the catalogue of Simard-Normandin
et al.\ and the joint one where it shifts to
$(l,b)=(110^\circ,41^\circ)$ and its intensity changes by 25 per
cent.  The position of the structure centred at
$(l,b)=(289^\circ,23^\circ)$ remains relatively stable
in all three catalogues. This feature is apparently connected with Loop~I and, possibly,
Loop~IV which can be seen at about the same position in radio continuum surveys
(Berkhuijsen, Haslam \& Salter 1971; Berkhuijsen 1971; Salter 1983).

The pattern of the signs of RM seen in Figs~\ref{Orion}b,c in
the general direction to the Galactic centre (with $\RM$ sign
changing as $-+-+$ when moving clockwise from the quadrant
$360^\circ>l>270^\circ,b>0$), was interpreted by Han et al.\ (1997)
as indicating an odd (dipolar), axisymmetric magnetic field in the
central region of the Milky Way.  However, the wavelet transform maps
presented here indicate that this pattern is just a part of a more
general RM distribution that extends wider along Galactic longitude
and is more plausibly due to local magnetic field distortions. The
significant north-south asymmetry is also better compatible with the
local nature of the $\RM$ features at these scales.

The available data on Faraday rotation measures of
extragalactic radio sources cannot provide stable and
selection-independent results for magnetic structures at angular
radii below $35^\circ$
when a global analysis of the kind performed here is applied.
This scale is equal to about seven times the typical separation of
data points in the catalogue of Simard-Normandin et al.\ (1981).  As
we show in Appendix~A, $20^\circ$ is the resolution limit for this
catalogue (and for that of Broten et al.\ 1988) associated with the
uneven distribution of the sources in the sky and their limited
number.  To check the robustness of our results, we
also considered the result of
removing sources with the largest $|\RM|$
in the catalogue of Simard-Normandin et al.;
the structures
at $R\simeq30$--$40^\circ$ remained stable (so that a structure does
not shift by more than its radius) even when 19 sources with
$|\RM|>200$ have been removed (of the total of 551 sources). Sources
with large positive $\RM$ occur mostly near $l\approx220^\circ$ and
those with large negative $\RM$ are concentrated near
$l\approx90^\circ$, therefore they are likely to trace real
structures of a small angular size. The signs of $\RM$ in these two
groups of sources are consistent with the direction of the local
large-scale magnetic field, although widely distributed sources with
moderate $\RM$ also provide a reliable, clear evidence of this field.
The results obtained from the joint catalogue are less stable even at
large scales, mainly because its data points are distributed less
evenly in the sky.
(We emphasize that the omission of sources with large $|\RM|$
was only used to assess the reliability of our results; our
conclusions reported below are all based on samples that do include
sources with large $|\RM|$.)

The resolution limit of the sample somewhat hampers analysis of
magnetic field features located far from the Sun because they span a
narrow range in $b$. Still, one can study some such details, provided
they are extended along longitude, by using only low-latitude sources
and performing a one-dimensional wavelet transformation in longitude.
We discuss such results in Sect.~\ref{LLS}.

Since the scale resolution of the wavelet family used here is only
moderate, we have restrained ourselves from discussing structures at
intermediate scales, but instead consider only two widely separated
scales, $76^\circ$ and $35^\circ$ both of which are distinguished by
maxima in the wavelet spectral energy density.

\setcounter{table}{0}
\begin{table*}
\label{res}
\caption{The dominant structures in the $\RM$ distribution}
\begin{center}
\begin{tabular}{@{}lcccccc@{}}
Name &$l_0$      &$b_0$      &$R$        &$W$         &$\B$       &$r_{\rm rev}$\\
&$(^\circ)$ &$(^\circ)$ &$(^\circ)$ &$(\!\radm)$ &($\mu$G) &(kpc)\\[3pt]
Sagittarius arm
        &$\phantom{2}33\pm5\phantom{2}$
        &--
                &$\phantom{<}35$
                        &$\phminus30\pm3\phantom{2}$
                                &$1.7\pm0.3$
                                        &\\
Carina arm
        &$330\pm5\phantom{2}$
        &--
                &$\phantom{<}35$
                        &$-30\pm8\phantom{2}$
                                &$1.6\pm0.3$
                                        &{\lower5pt\hbox{7.9}}\\
Orion arm
        &$\phantom{2}98\pm18$
        &$-13\pm17$
                &$\phantom{<}76$
                        &$-27\pm4\phantom{2}$
                                &$1.5\pm0.3$
                                        &\\
Orion arm
        &$276\pm17$
        &$-17\pm9\phantom{1}$
                &$\phantom{<}76$
                        &$\phminus24\pm8\phantom{2}$
                                &$1.4\pm0.3$
                                        &{\lower5pt\hbox{10.5}}\\
Perseus arm
        &$130\pm30$
        &--
                &$<30$
                        &$\phminus20\pm10$
                                &$1.5\pm1.2$
                                        &\\
Perseus arm
        &$190\pm5\phantom{2}$
        &--
                &$<30$
                        &$-30\pm10$
                                &$1.2\pm0.7$
                                        &\\[3pt]
Loop~I
        &$289\pm16$
        &$\phminus23\pm10$
                &$\phantom{<}41$
                        &$-28\pm8\phantom{2}$
                                &$0.9\pm0.3$
                                        &\\[4pt]
\multicolumn{7}{@{}p{107mm}@{}}
{\footnotesize
{\bf Notes:} $l_0$ and $b_0$ are the Galactic coordinates of the
centre of a structure, $R$ is its angular radius (scale), and
$W=w(R,l_0,b_0)$ is its amplitude; $B$ is the
total strength of the mean magnetic field,
except for Loop~I where the line-of-sight component is given;
$r_{\rm rev}$ is the galactocentric radius of a  magnetic field
reversal between the corresponding arms at $l=0$ assuming the radius
of the Solar orbit of $8.5\kpc$. Negative (positive) values of $W$
indicate that the field is directed toward (away from) the observer.
Results for the
Sagittarius arm (and its extension to $360^\circ>l>270^\circ$,
the Carina arm) and the Perseus arm have been obtained from
low-latitude sources, $|b|\leq10^\circ$.  }
\end{tabular}
\end{center}
\end{table*}


To conclude this section, we present in Table~1 parameters of the
stable structures discussed in Sect.~\ref{MFOA} where magnetic
field strengths also are estimated.
The mean values of the parameters for the Orion
arm and Loop~I have been obtained using the catalogue of
Simard-Normandin et al.\ (1981).  Their errors (except for magnetic
field strength) were obtained as a maximum deviation, from the above
mean value, of the values obtained from the joint catalogue and
pairwise combinations of the catalogue of Simard-Normandin et al.\
(1981) with those of Tabara \& Inoue (1980) and Eichendorf \&
Reinhardt (1980).  When estimating the errors, we omitted sources
with $|b|<10^\circ$ from the latter two catalogues and the joint
catalogue. Furthermore, we added squares of the above maximum
deviations and of the average angular separation of the measured
points in the catalogue of Simard-Normandin et al., equal to
$4^\circ$, to obtain the final errors of $l_0$ and $b_0$.  In a
similar manner, the error of $W$ given includes the uncertainty of
the normalization constant $c=h-\Ko$ (see Eqs~(\ref{psi_tild}) and
(\ref{C})), which is equal to $2\radm$ for the catalogue of
Simard-Normandin et al. Individual features are further discussed in
Sect.~\ref{MFOA}. The feature labelled Loop~I in Table~1 occupies
only a part of Loop~I as described by Berkhuijsen (1971); in
particular, the brightest part of the radio Loop~I, the North Polar
Spur, is slightly shifted from the above feature of the $\RM$
distribution.

The results obtained from the catalogue of Broten et al.\ agree with
those from the data of Simard-Normandin et al.\ to within 1/3 of the
errors quoted in Table~1.

\section{The large-scale magnetic field}       \label{MFOA}
In this section we discuss the magnetic field in the Galactic spiral
arms and in Loop~I considering separately high- and low-latitude
sources.  We estimate magnetic field strength in the Orion
arm and Loop~I in Sect.~\ref{SS} and then consider low-latitude
sources in Sect.~\ref{LLS} to study magneto-ionic structures in the
Sagittarius, Carina and Perseus arms.

\subsection{The Orion arm and Loop~I}       \label{SS}
In order to estimate the strength of the large-scale magnetic field
in the Orion arm we compared the wavelet transform for $\RM$ in the
whole range of $b$ from the catalogue of Simard-Normandin et al.\
(1981) with those for  model magnetic field distributions and
adjusted $\B$ to obtain, at $R=76^\circ$, structures of the same
intensity $w$ and localization $(l_0, b_0)$ as in the observed $\RM$
map. It is important to stress that the model wavelet transforms were
always calculated on the real data grid (551 points positioned as in
the catalogue of Simard-Normandin et al.). We have verified that the
results change insignificantly when sources within the belts
$|b|<5^\circ,10^\circ$ and $15^\circ$ are omitted.

The model magnetic field was purely azimuthal with strength
independent of both galactocentric radius and the vertical
coordinate. The half-thickness of the magnetic field distribution was
adopted as $h=1000\p$ (twice the scale height of the warm
interstellar gas) and its radius as $R\g=15\kpc$. We considered
models with uniform thermal electron density and also with the
electron density distribution of Cordes et al.\ (1991, see also
Taylor \& Cordes 1993),
\begin{eqnarray}
\nel&=&n_1+n_2\;,                       \label{cordes}\\
\frac{n_1}{1\cmcube}&=&0.025\exp\left[-{{|z|}\over{1\kpc}} -\left({r\over
{20\kpc}}\right)^2\right], \label{cordes1}\\
\frac{n_2}{1\cmcube}&=&0.2\exp\left[-{{|z|}\over{0.15\kpc}}
-\left({{r-4\kpc}\over{2\kpc}}\right)^2\right].
        \label{cordes2}
\end{eqnarray}
The radius of the solar orbit was adopted as $R\solar=8.5\kpc$. The
term $n_1$ describes a smooth disc, and $n_2$ represents a
dense annulus at $r=4\kpc$. Note that Heiles, Reach \& Koo (1996)
argue that this annulus does not exist. We also tried $h=2000\p$ and
$R\g=20\kpc$ for the size of the magnetic disc to check that our
results are affected only insignificantly in models with $\nel$ given
by Eqs~(\ref{cordes})--(\ref{cordes2}).
We consider a purely azimuthal magnetic field in these fits
because the quality of the data does not warrant more complicated
models with non-zero radial field. The data for low-latitude sources
discussed in Sect.~\ref{OG} do not admit such a simple model, and so
allow us to estimate reliably the magnetic pitch angle.

We calculated model $\RM$ for all 551 points and then performed their
wavelet transform. For a uniform disc model the results are as
follows. The maxima of $|w|$ occur at the same scale $R=76^\circ$ as
for the real data.  The value of magnetic field was found by
adjusting the value of the model wavelet transform $w$ at the extrema
to those for the catalogue of Simard-Normandin et al.;
$B\approx1.2\mkG$ fits both the extrema.  The main deficiency of this
model is that the extrema of the wavelet transform are displaced from
the observed structures towards $l=0$ by about $20^\circ$.

Therefore, we further tried a nonuniform electron density
distribution retaining only $n_1$, Eq.~(\ref{cordes1}). Then the
positions of the extrema agree much better and, estimating $B$ as
above, we obtain $1.4$ and $1.5\mkG$ for the positive and negative
extremum, respectively; these values are cited in Table~1. The
difference of the values of $B$ obtained in various models gives the
uncertainty of $B$ as shown in the table.

The annulus described by $n_2$ in Eq.~(\ref{cordes2}) is projected
onto a small region at $|b|\leq10^\circ$ and so it does not affect
rotation measures at large scales.

\epsfxsize=7.9cm
\begin{figure}
\centerline{\epsfbox{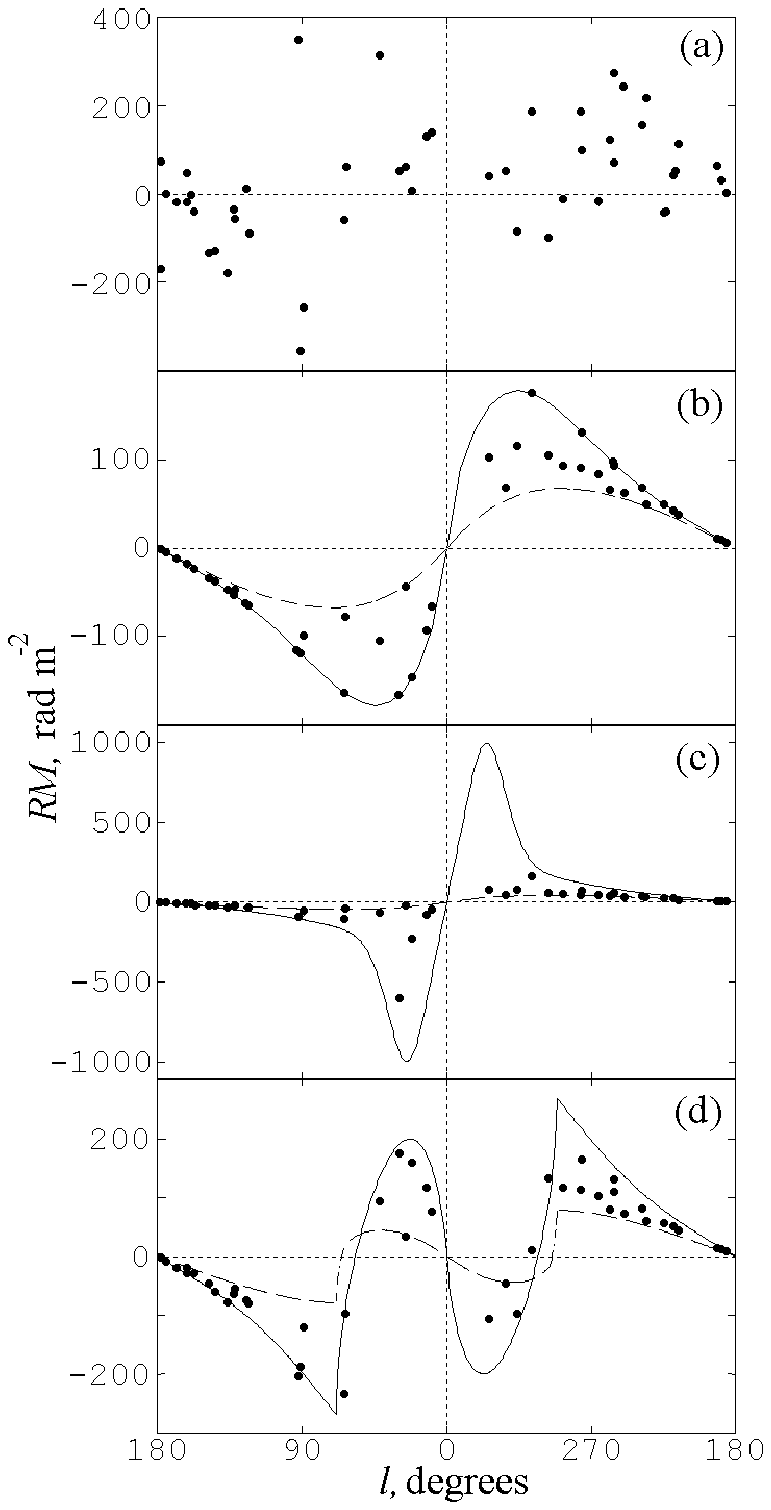}}
\caption[]{The longitudinal distributions of $\RM$ for low-latitude
sources, $|b|\leq10^\circ$, (a) in the catalogue of Simard-Normandin
et al.\ (1981) and for the following four models: (b) a uniform
azimuthal magnetic field and $\nel=n_1$, see
Eqs~(\protect\ref{cordes})--(\protect\ref{cordes2}), (c) a uniform
azimuthal field and $\nel=n_1+n_2$, and (d) an azimuthal field with a
reversal at $r=7.9\kpc$ in a disc with $\nel=n_1$. The model $\RM$
distribution at $b=0$  is shown with solid line and that at
$b=10^\circ$ is shown dashed; dots show the values of $\RM$, as
observed (panel a) and calculated for the true positions of the
sources (panels b--d) using the model magnetic field.}
\label{low-lat}
\end{figure}
\begin{figure}
\epsfxsize=8.3cm
\centerline{\epsfbox{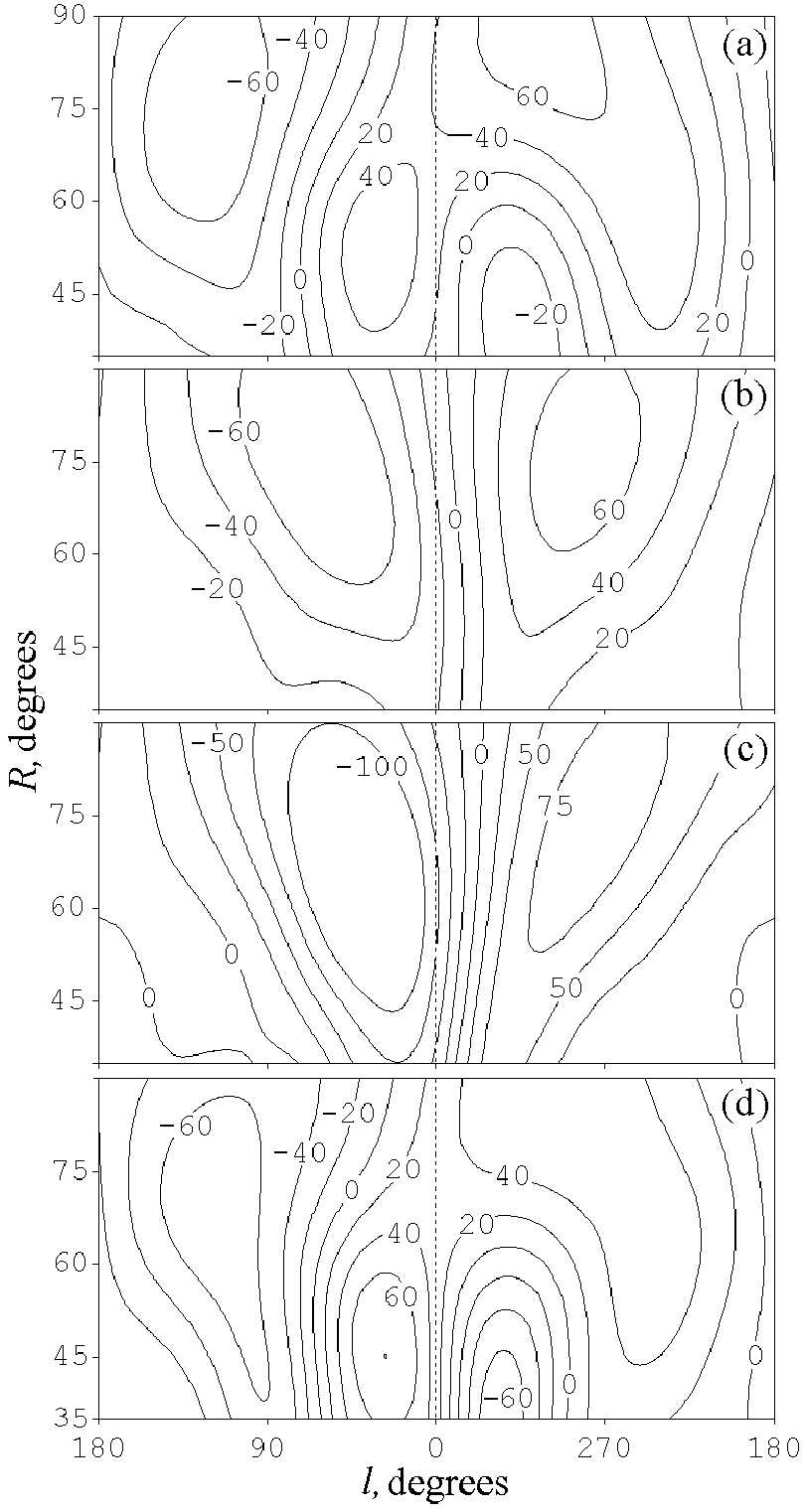}}
\caption[]{One-dimensional wavelet transforms of the distributions
shown in the respective panels of Fig.~\protect\ref{low-lat} shown in
the $(l,R)$ plane with $R$ the scale of the structure. Contour labels
are given in $\!\radm$.}
\label{oneD}
\end{figure}

The average line-of-sight magnetic field in Loop~I
can be estimated as $0.9\pm0.3\mkG$ from the contribution of Loop~I
to the Faraday rotation measure given in Table~1, $|W|=28\pm8\radm$,
using the ionized gas density in the shell of $0.4\cmcube$ and
pathlength of $100\p$ (Heiles
et al.\ 1980, Salter 1983). The amplitude of the negative extremum
identified with Loop~I, $W=-28\pm8\radm$ is close to the estimate of
Faraday rotation measure in the North Polar Spur, $\simeq+26\radm$
(Heiles et al.\ 1980), but the signs are different. The latter
estimate refers to a dense shell of Loop~I at
$(l,b)=(32^\circ,42^\circ)$, whereas the former to its part on the other
side of the shell centre. The ensuing geometry of magnetic field in the
shell of Loop~I is compatible with interstellar horizontal magnetic field bent into an
arc and compressed by expanding shell, so that the
line-of-sight magnetic field has opposite directions
on the opposite sides of the shell centre. The North Polar Spur
itself has a scale smaller than those discussed in this paper, but
the structure of positive $\RM$ at $(l,b)=(50^\circ,21^\circ)$ in
Fig.~5b can be related to it.  Our estimate of magnetic field in
Loop~I agrees with the results of Troland \& Heiles (1982) who argue
in favour of a weak magnetic field enhancement in the shell of
Loop~I.

The $\RM$ features attributed to the reversed magnetic field of
the Sagittarius arm occur at about the same Galactic longitudes (and
have the same signs) as Loop~I and its most prominent part, the North
Polar Spur. The connection of these features with a remote region is
confirmed by the studies of pulsar Faraday rotation measures mentioned
in Sect.~\ref{SRMS}. It is also noteworthy that the North
Polar Spur is not well pronounced at low latitudes $b<8^\circ$
(Berkhuijsen et al.\ 1971) where the structure attributed to the
Sagittarius arm is well visible.

To conclude this section, we give a crude estimate of the size of the
region sampled by the extragalactic sources. Even though all sources
have been taken into account to estimate $B$, the main contribution
to $\RM$ (and the wavelet transform) is due to a region within a
distance $L_B=W/(K \bar n_{\rm e} B \cos \alpha )$, where $\bar
n_{\rm e} = 0.03\cmcube$ is the typical value of electron density and
$\alpha$ is the average angle between the magnetic field direction
(assumed to be from $l=270^\circ$ to $l=90^\circ$) and the direction
to a source. This yields $2L_B\simeq3\kpc$ for the size (along the
field direction) of the region sampled.  The low-latitude sources
sample a more extended region.

\subsection{Low-latitude sources and magnetic field reversals}
                                                                \label{LLS}
As discussed in Sect.~\ref{SIFRS}, the resolution limit of the
catalogues used here is about $20^\circ$.  Therefore, features at
$|b|\la 10^\circ$ cannot be studied using an isotropic
two-dimensional wavelet.  Hence, we considered low-latitude sources
separately to study remote structures located outside the Orion arm
whose radius is larger than $20^\circ$ in $l$, but smaller than
$10^\circ$ in $b$.

The large-scale magnetic field in the Orion arm produces rather
different structures in the $\RM$ distributions at low and high
Galactic latitudes. Sources at latitudes $|b|>b_*$ sample magnetic
field in a region around the Sun whose radius is $L<h/\sin{b_*}$.
Under a crude assumption of a purely azimuthal field, the curvature
of the field lines of the large-scale magnetic field can be neglected
provided $L<R\solar$, or $|b|>7^\circ$ for $h=1\kpc$ and
$R\solar=8.5\kpc$. For sources at lower Galactic latitudes,
$|b|<b_*$, the curvature of the magnetic lines cannot be neglected.
We adopt the borderline latitude as $b_*=10^\circ$.

Hence, maximum values of $|\RM|$
produced in the Orion arm should be observed at
$l\approx90^\circ$ and $270^\circ$ for sources at $|b|\ge 10^\circ$ even for
curved magnetic lines, but for low-latitude sources the largest
$|\RM|$ are shifted towards $l=0$.
Therefore, high- and low-latitude sources should be considered
separately.

\subsubsection{The catalogue of Simard-Normandin et al.}
\label{IG}
The catalogue of Simard-Normandin et al.\ (1981) contains 48 sources
with $|b|\leq10^\circ$ which are shown in Fig.~\ref{low-lat}a.  The
scatter of the points is explained, in part, by scatter in the
latitudes of the sources. We show in Figs~\ref{low-lat}b--d the
values of $\RM$ for the same 48 positions, but  calculated from
three model distribution of magnetic field $B$ and electron density
$\nel$ described in what follows.  In order to provide a measure of
the variations of $\RM$ with $b$, we present in these figures the
model values of $\RM$ at both $b=0$ (solid) and $|b|=10^\circ$
(dashed). Note that the maximum values of $|\RM|$ may occur away from
$b=0$ if the field has a reversal because the maximum value can be
attained on the line of sight with $|b|\neq0$ that passes through the
Orion arm but not through a more distant region with reversed
magnetic field.

The random Galactic magnetic
field leads to a random walk of $\RM$, and the associated random
scatter of $\RM$ is especially strong for low-latitude sources where
the number of turbulent cells along the line of sight is large. The
data are further contaminated by the intrinsic rotation measures of
the radio sources. To filter out the random fluctuations that
dominate at small scales, we compare with each other the wavelet
transforms of the observational data and the model distributions,
shown in Fig.~\ref{oneD}, rather than the original distributions of
$\RM$ along $l$ shown in Fig.~\ref{low-lat}.

We performed one-dimensional wavelet transformation (along $l$) for
the data in Fig.~\ref{low-lat} and show in Fig.~\ref{oneD} the
wavelet transforms in the $(l,R)$ plane. The one-dimensional `Mexican
hat' was used as the analyzing wavelet.  Unlike maps in the $(l,b)$
plane discussed above, Fig.~\ref{oneD} conveniently displays all the
scales simultaneously. Figure~\ref{oneD}a shows the wavelet transform
for the data of Simard-Normandin et al., Fig.~\ref{low-lat}a. The
following three model distributions are presented in
Figs~\ref{low-lat}b--d and their wavelet transforms are shown in the
corresponding panels of Fig.~\ref{oneD}.  The first model
(Figs~\ref{low-lat}b and \ref{oneD}b) is a disc of a radius $15\kpc$
with a uniform azimuthal magnetic field of a strength $B$ (an
adjustable parameter) and the electron density  given by
Eq.~(\ref{cordes1}).  The positions of maximum $|\RM|$ in this model
are strongly shifted towards $l=0$ and occur at $l\approx40^\circ$
and $320^\circ$.  The asymmetry between the left- and right-hand
sides of these and other similar plots is entirely due to asymmetry
in the data point distribution where more sources at $0<l<180^\circ$
happen to probe regions with large $|\RM|$. We evaluated $B$ by
minimizing the r.m.s.\ deviation of $w$ from the observed values of
Fig.~\ref{oneD}a to obtain $B=0.8\mkG$. These results change
insignificantly if we take $\nel = 0.03 \cmcube$ independently of
radius.

The second model (Figs~\ref{low-lat}c and \ref{oneD}c) has a uniform
azimuthal magnetic field, but the electron density contains both
terms of Eq.~(\ref{cordes}). This model results in very strong
extrema of $\RM$ located at $l\approx25^\circ$ and $335^\circ$. Only
a few sources in the sample are projected on the annulus $n_2$
because it occupies a narrow range $|b|\la1^\circ$ at
$l\approx25^\circ$ and $335^\circ$.

The above models produce only two clearly visible maxima of $\RM$,
whereas the observed picture, Fig.~\ref{oneD}a, has a more
complicated structure with four local extrema.  Therefore, we had to
consider a more complicated model that uses again $\nel=n_1$, but
contains a reversal of azimuthal magnetic field at a certain
galactocentric radius $r = r_{\rm rev}$ (the results shown in the
figures correspond to $r_{\rm rev}=7.9\kpc$).  The wavelet transform
in Fig.~\ref{oneD}d contains an additional pair of structures
produced by the reversed magnetic field.  The model of
Fig.~\ref{oneD}d clearly provides a fair fit to the data of
Fig.~\ref{oneD}a.  We adjusted both $B$ (assumed to be uniform apart
from the reversal) and $r_{\rm rev}$ to reach the best agreement of
the model and the `observed' values of $w$ by minimizing the r.m.s.\
deviation. A minimum deviation is achieved for $r_{\rm rev}=7.5\kpc$
and $B=1.4\mkG$.  However, the best overall agreement in terms of the
shapes and positions of all the structures occurs for $r_{\rm
rev}=7.9\kpc$ and $B=1.3\mkG$.  Both estimates of $B$ in the Orion
arm agree within errors with that from $\RM$s of high-latitude
sources.  Magnetic field in the Sagittarius and Carina arms is
estimated in Sect.~\ref{OG}.

A notable feature of Fig.~\ref{oneD}d is that the field reversal in
the inner Galaxy can be clearly seen in the wavelet transforms of
both the original and model data in both first and fourth Galactic
quadrants. There are only a few sources with negative $\RM$ in the
Carina arm region (the fourth quadrant), so that one cannot isolate
the reversal there using the raw data of Figs~\ref{low-lat}a,d in the
$(\RM, l)$ plane (Simard-Normandin \& Kronberg 1979, 1980).  Wavelet
transformation provides efficient scale filtering and thereby allows
us to isolate weaker localized details seen against the background of
an extended and stronger signal, and so the reversed field can be
easily identified at both $0<l<90^\circ$ (the Sagittarius arm) and
$270^\circ<l<360^\circ$ (the Carina arm).

We note that the model with reversal presented in
Fig.~\ref{low-lat} reproduces well $\RM$ in the Cetus arc region.
This is consistent with the fact that Loop~II cannot be distinguished
at low Galactic latitudes (Berkhuijsen 1971).

Our fits to the data of Simard-Normandin et al.\  have a
purely azimuthal magnetic field. We stress that this does not imply that
the magnetic field in the Milky Way is purely azimuthal in reality,
but merely indicates that its radial component cannot be estimated reliably from
these data. Therefore, we prefer here a model with the minimum number of parameters; the
data available do not warrant using more complicated (and realistic) models.  For the
same reason we do not employ a model of electron density including spiral arms (Taylor \&
Cordes 1993). The data of Broten et al.\ (1988) used in Sect.~\ref{OG} are not compatible
with a purely azimuthal field, so we apply a more complicated model and estimate the
pitch angle of magnetic field there.

\subsubsection{The catalogue of Broten et al.}       \label{OG}
The catalogue of Broten et al.\ (1988) contains 69 low-latitude
sources (21 more than the catalogue of Simard-Normandin et al.).
There are 7 low-latitude sources in Simard-Normandin et al.\ which
are absent in Broten et al.\ and which have also been included in the
sample used in this section. A one-dimensional wavelet transform of
these data (76 sources shown in Fig.~9a) is shown in Fig.~8a in the
same format as in Fig.~\ref{oneD}a.  The large-scale structures due
to the Orion arm, centred at $l\approx90^\circ$ and
$280^\circ$, are similar to those obtained from the data of
Simard-Normandin et al.  This is true also for the reversed magnetic
field in the Sagittarius--Carina arm that produces structures at a
scale $35^\circ$--$45^\circ$, slightly larger than
from the high-latitude sources.

A remarkable feature of Fig.~8a is that it has a new
pair of structures centred at a scale less than $30^\circ$. This is a
positive extremum at $l\approx125^\circ$ seen against a negative
background produced by the local magnetic field in the Orion arm and
a negative extremum at $l\approx190^\circ$. If these structures can
be attributed to the Perseus arm, magnetic field in that arm must be
directed oppositely to the local field, in agreement with the earlier
determination of Agafonov et al.\ (1988).
The Perseus arm is seen at $l=145^\circ$--$160^\circ$ in
synchrotron intensity and at $l=90^\circ$--$140^\circ$ in the optical range (Berkhuijsen
1971). The negative extremum attributed to the Perseus arm occurs close to the anticentre
direction because this relatively weak feature is best visible in the direction where the
local contribution from the Orion arm is minimum.
(Magnetic field reversal in the outer Galaxy was also advocated by Rand \& Kulkarni
1989 and Clegg et al.\ 1992.)

The sample of 76 low-latitude sources includes some with very large
values of $|\RM|$ (with the maximum of $800\radm$).  To assess the
robustness of our results we removed sources with the largest $|\RM|$
reducing gradually the cut-off value. We show in Figs~9b and 8b
the the original data and their wavelet transform for 68 sources with
$|\RM|<300\radm$.  The newly detected pair of structures becomes less
pronounced, but remains well visible.

The simplest model which is able to produce a similar pair of
structures contains an additional reversal of magnetic field in the
outer Galaxy. We first tried purely azimuthal field as in
Sect.~\ref{IG}, but this failed to produce the required structures of
the desired strength and location for any realistic field strength.

\epsfxsize=8.2cm
\begin{figure}
\label{brot_fig}
\centerline{\epsfbox{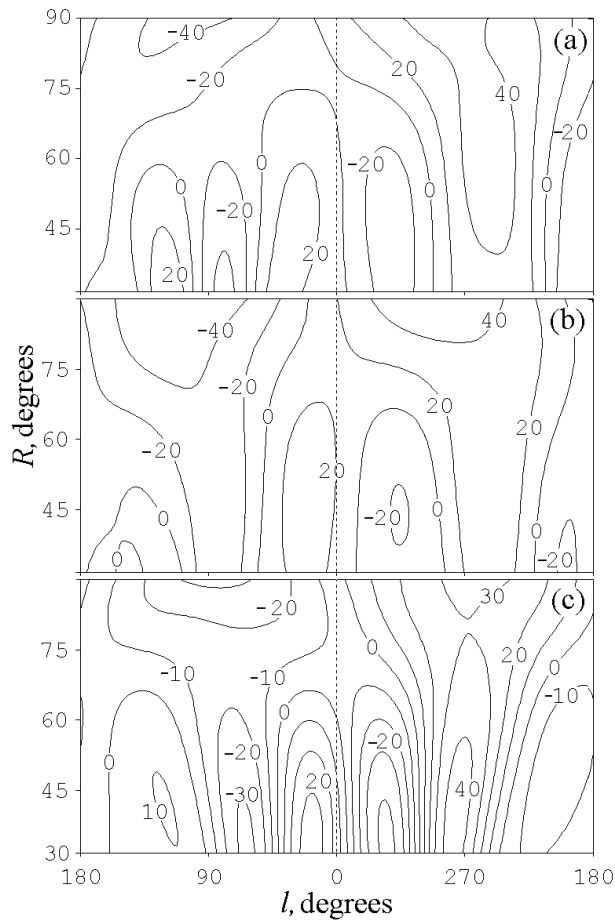}}
\caption[]{One-dimensional wavelet transforms for the combined set of
76 low-latitude sources, $|b|\leq10^\circ$,  from Broten et al.\
(1988) and Simard-Normandin et al.\ (1981): (a)  all the sources; (b)
those with $|\RM|<300^\circ$; and (c) model with two reversals
and $\nel=n_1$ as described in the text. Contour labels are in
$\!\radm$.}

\end{figure}
\begin{figure}
\epsfxsize=8cm
\centerline{\epsfbox{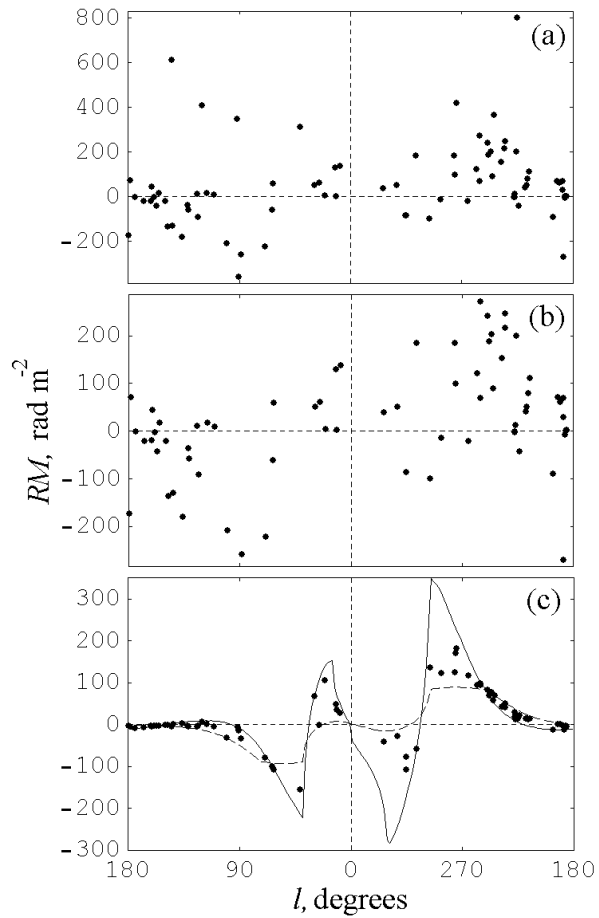}}
\caption[]{The $\RM$ data whose  wavelet transforms are shown in the
respective panels of Fig.~8. Curves shown in (c)
have the same meaning as in Fig.~\protect\ref{low-lat}.
}
\label{rm3arm}
\end{figure}

Therefore, we had to consider a more complicated model where magnetic
lines are trailing logarithmic spirals. The simplest model of this kind
includes three logarithmic spiral branches (for the Sagittarius--Carina,
Orion and Perseus arms) with the pitch angle $p=15^\circ$ and with
magnetic field reversals between (or within) each arm. The value of
the pitch angle adopted is close to $14^\circ$ as inferred for the Galactic
spiral arms from synchrotron intensity observations by Beuermann et al.\
(1985). With this model, we adjusted the strength of the
magnetic field in each arm and the positions of the reversals. The
best fit between the wavelet transforms for the original and model RM
is obtained for the parameters given in Table~1 (for the Orion arm,
$B\approx1.5\mkG$ was obtained).  The wavelet transform of the model
RM is shown in Fig.~8c; we also show both the original
and model RMs in Fig.~9 in the same format as in
Fig.~\ref{low-lat}. The structures due to the Perseus arm occur at
$90^\circ\leq l\leq 200^\circ$, where the sample of Broten et
al.\ (1988) contains 13 additional sources. The errors in $W$ and
$B$ given in Table~1 characterize the scatter of the values obtained
from fitting to the data of Simard-Normandin et al.\
(Fig.~\ref{oneD}a) and Broten et al.\ (Figs~8a,b).

\epsfxsize=6.5cm
\begin{figure}
\vspace*{1.7cm}
\centerline{\epsfbox{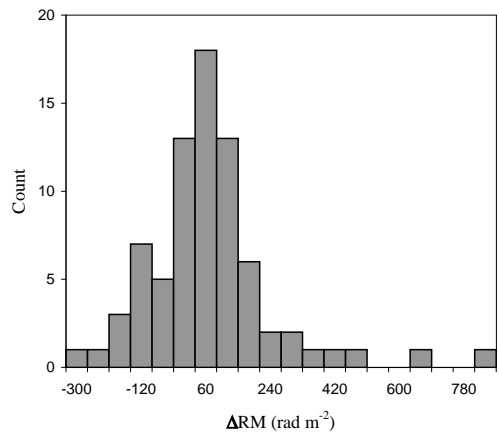}}
\caption[]{The histogram of the difference between the observed
$\RM$ of the low-latitude sources of Broten et al.\ (1988) and
Simard-Normandin et al.\ (1981) and the best-field model $\RM$ with
two magnetic field reversals; the bin width is $60\radm$.}
\label{histogram}
\end{figure}

The histogram of $\Delta\RM$, the difference between the
observed and model values of $\RM$ for the best fit model, is shown
in Fig.~\ref{histogram}. The histogram is as close to a Gaussian as can be expected with
76 data points; the tail at positive values is due to sources with $|\RM|>300\radm$. The
mean value of $\Delta\RM$ in Fig.~\ref{histogram} is $40\radm$ for all 76 sources, but
just $4\radm$ for sources with $|\RM|<300\radm$.  The standard deviation of $\Delta\RM$
is $\sigma\approx100\radm$ for the sources with $|\RM|<300\radm$ (and $180\radm$ if the
eight sources with $|\RM|>300\radm$ are included). A similar amount of scatter is implied
by the observed rotation measure structure function, ranging from about $40\radm$ at a
high latitude $b=-20^\circ$ (Minter \& Spangler 1996) to $200\radm$ at low latitudes
($0<b<4^\circ$) (Lazio, Spangler \& Cordes 1990) at a scale $30^\circ$. A rough estimate
of the scatter expected to arise within the Milky Way is $\sigma\simeq K\nel
B[(q_B^2+q_n^2)dL]^{1/2}\simeq40\radm(q_B^2+q_n^2)^{1/2}$, where
$q_B=\delta B/B$ and $q_n=\delta\nel/\nel$ are the relative
fluctuations in magnetic field and electron density, $d\simeq100\p$
is the turbulent scale and $L\simeq3\kpc$ is the pathlength (see
Minter \& Spangler 1996 for details).

We note that the above fits including the Perseus arm place the
reversal in the inner Galaxy at $r=6.5\kpc$, i.e., somewhat farther
from the Sun than the model of Sect.~\ref{IG}.  We consider the model
of Sect.~\ref{IG} to be more accurate in this respect and the
corresponding estimate of the reversal radius is given in Table~1.

The Orion arm can be different from the other arms being a short
spur rather than a long arm; then its pitch angle can be
significantly different from that of the Sagittarius and Perseus
arms. If this is the case, the fit of this section will be
distorted by our equal treatment of the three arms.
Further refinement of the model will be useful when more extensive
RM catalogues will become available. The presents fits look good
enough without such refinements, so more complicated models cannot be
justified.

We do not see any signatures of the Scutum arm, an inner spiral arm
next to the Sagittarius arm. Its longitudinal range is $24^\circ\la
l\la43^\circ$ (e.g., Beuermann et al.\ 1985,  Vall\'ee et al.\ 1988). Assuming the scale
height of the regular magnetic field to be $1\kpc$ and the arm
galactocentric distance of $5\kpc$ at $l=0$, the distance to the
tangential point is more than $6.3\kpc$; thus, the Scutum arm spans
the latitude range of $|b|\la9^\circ$. This window contains just two
sources in the catalogue of Broten et al.\ (1988), one with
$\RM=50\radm$ and the other with $\RM=-840\radm$, with the latter
value of $|\RM|$ being too large to be attributed to the Galaxy.
Vall\'ee et al.\ (1988) claim that the Scutum arm produces a negative
contribution to $\RM$ using 9 sources 7 of which have negative
$\RM$. However, 6 of their sources with $\RM<0$ have
$|b|>10.4^\circ$.  Moreover, two of them have $\RM=-699$ and
$-840\radm$; such values of $\RM$ cannot be produced in the Milky
Way. Vall\'ee et al.\ justify their wider window in latitude,
$|b|<14^\circ$, by taking an overestimated magnetic field scale
height of $1.25\kpc$ and underestimated distance to the tangent point
($5\kpc$). The scale height of the regular magnetic field is about
$1\kpc$ in the disc near the Sun (e.g., Fletcher \& Shukurov 2000)
and even smaller in the inner Galaxy (Phillipps et al.\ 1981, Beuermann et al.\ 1985).

\subsection{The vertical symmetry of the large-scale magnetic field}\label{VS}
Although the structures in the $\RM$ distribution at the largest
scale are slightly shifted to negative Galactic latitudes, they
clearly indicate that the large-scale magnetic field in the Orion arm
has the same direction above and below the Galactic midplane (see
Sect.~\ref{SIFRS}).  This result agrees with most earlier studies of
the Galactic magnetic field (with a few exceptions -- see
Sect.~\ref{SRMS}). However, earlier results were based either on a
naked-eye inspection of noisy $\RM$ maps or on fitting magnetic field
models having presupposed even symmetry.  Our results
provide a rare objective confirmation of the anticipated even
symmetry of the horizontal magnetic field in the Orion arm of the
Milky Way.

The fact that the field direction is the same above and below the
Galactic equator is compatible with the overall quadrupole symmetry
of magnetic fields generated by disc dynamos. However, this cannot be
considered as a completely convincing proof of the dynamo
origin of the Galactic magnetic field because the dynamo-generated
quadrupole magnetic field must also have a vertical magnetic field
antisymmetric with respect to the midplane.  Hence, strictly
speaking, one has to measure $\B_z$ before it can be concluded
definitely that the observed symmetry directly confirms galactic
dynamo models. Still, the even symmetry of the azimuthal field proved
here significantly restricts theories of magnetic field origin (see
Beck et al.\ 1996).

Even when the contribution of Loop~I has been subtracted, the centres
of the dominant $|\RM|$ maxima for high-latitude sources remain
shifted from the midplane by about $15^\circ$ to the south.  (We have
verified that the overall symmetry remains even when we
discard also the data in a region in the Southern hemisphere located
symmetrically to Loop~I.) This residual asymmetry can hardly be
explained by the shift of the Sun's position away from the Galactic
midplane.  A displacement of the Sun by about 400\,pc to the north
would be required to reproduce the observed residual asymmetry,
whereas the actual displacement is as small as 50\,pc.

Another cause of the asymmetry could be a local contamination
from Loop~II, centred at $(l,b)=(110^\circ,-32^\circ)$, whose radius
is $50^\circ$ (Berkhuijsen 1971). However, there are no local
magneto-ionic structures of a similar size located symmetrically to
Loop~II in the fourth quadrant (the Gum nebula has a small angular radius of
$18^\circ$ and occurs at $b=0$ -- e.g., Vall\'ee 1993).  Therefore,
the symmetry of the residual shift in the large-scale wavelet extrema
between the first and fourth Galactic quadrants does not support
this idea.

The residual shift of the centres of the dominant $\RM$ structures to
negative Galactic latitudes can be a result of a genuine global
vertical asymmetry of the magnetic field with a stronger large-scale
field in the southern hemisphere. The possibility of an asymmetry of
this type was suggested by Sokoloff \& Shukurov (1990) as a
consequence of the odd symmetry of magnetic field in the Galactic
halo (see also Brandenburg et al.\ 1992; Brandenburg, Moss \&
Shukurov 1995). The odd symmetry of the halo magnetic field has been
suggested by the spherical harmonic analysis of extragalactic $\RM$
by Seymour (1984).  The shift to $b_0\simeq-15^\circ$ then implies
that the vertical distribution of the field has a maximum at
$z=-L_B\tan{b_0}\simeq400\p$. We stress that this estimate is only
illustrative and has an uncertainty of at least 50 percent.

This interpretation is not unique. A similar asymmetry would arise
due to a local perturbation in magnetic field and/or
electron density. We can estimate the implied properties of such a
perturbation assuming, for example, that the Sun is located near the
top of a magnetic loop extending vertically above the midplane. Such
a loop can result from the Parker instability.  If the base of the
loop is at the midplane, $z=0$, and the Sun is at the height
$z\solar=50\p$, then the horizontal size of the loop is $L\simeq
2z\solar/|\tan{b_0}|\simeq400\p$. The difference between the values
of $w$ at $b=-15^\circ$ and $b=+15^\circ$ for $l=98^\circ$ in
Fig.~\ref{Orion}a is $\Delta w\simeq4\radm$. Then the field
enhancement in the loop (relative to the average field) is estimated
as $\Delta B \simeq \Delta w/(K\nel L)\simeq0.5\mkG$.

\section{Conclusions}                    \label{D}

Using the technique of wavelet analysis adapted to nonuniform data
grids, we have studied the large-scale Galactic magnetic field using
Faraday rotation measures of extragalactic radio sources. Our main
results are based on the catalogues of Simard-Normandin et al.\
(1981) and Broten et al.\ (1988). We have also incorporated some
measurements from other catalogues into a joint catalogue containing
841 source.

We have tested the stability of the results with respect to
modifications of the data sample. In particular, results obtained
from the joint catalogue are rather unstable confirming that the
uniformity of the data sample is important. We use other catalogues
to select unreliable measurements in the catalogue of
Simard-Normandin et al.\ (1981) and discard only four sources (see
Sect.~\ref{DDS}); this agrees well with the value expected from
statistical arguments (P.P.~Kronberg, private communication).
Therefore we used only 551 sources from the catalogue of
Simard-Normandin et al.\ (1981).

The large-scale magnetic field in the local spiral arm is dominated
by the azimuthal component. We confirm that there is a reversal in
the magnetic field in the inner Galaxy at a distance of 0.6--$1\kpc$
from the Sun.  Using an axisymmetric model of the Galactic
distribution of thermal electrons (Cordes et al.\ 1991), we estimate
the strength of the magnetic field in the Orion arm as
$1.4\pm0.3\mkG$.

The extension of the magnetic fields reversal between the Orion and
Sagittarius arms to the Carina arm region at negative Galactic
latitudes has been detected.  Faraday rotation measures of
low-latitude sources from the catalogue of Broten et al.\ (1988) are
consistent with a field reversal between the Orion and Perseus arms.
The average local pitch angle of the magnetic field is about
$15^\circ$.

The extent, in the azimuthal direction, of the region with
reversed large-scale magnetic field in the Solar vicinity cannot be
determined from the available data. We can only set its lower limit
at $3\kpc$. Therefore, the region can be relatively small not only in
radius (2--3\,kpc) but also in azimuth, rather than extended over the
whole circle around the Galactic centre (if the field is
axisymmetric) or along a spiral line (for a bisymmetric field)
(Shukurov 2000). Morphologically, this would be in accord with the
Orion arm being a short spur between the Sagittarius and Perseus arms
-- the field would be reversed in this spur. A similar magnetic
structure has been found in the galaxy M51 by Berkhuijsen et al.\
(1997), where  the regular magnetic fields in the disk is reversed in a
region of a size about 8\,kpc in azimuth and 3\,kpc in radius.

The dominant even symmetry of the large-scale local magnetic
field with respect to the Galactic equator has been demonstrated
convincingly using wavelets. This symmetry becomes obvious after filtering
out $\RM$ structures at angular scales smaller than about $70^\circ$
(these smaller-scale structures include Loop~I).

Our results are consistent with a moderate asymmetry of the local
large-scale Galactic magnetic field with respect to the equator (with
the $|\RM|$ maxima occurring at about $b=-15^\circ$) that can be
explained by stronger field in the southern Galactic hemisphere or by
the Sun being located close to the top of a magnetic loop such as
that arising from the Parker instability.  The horizontal size of the
asymmetric region is at least $3\kpc$ if it is global and $400\p$ if
it is local.

Loop~I affects the values of $\RM$ within a region which
extends to $l\approx50^\circ$, that is by $10^\circ$--$20^\circ$ farther than the
edge of Loop~I as seen in synchrotron emission.

\appendix
\section{Wavelet analysis for discrete
data irregularly distributed on a sphere }
A basic property of wavelets is that their average value vanishes,
see Eq.~(\ref{w-adm}).  In an ideal situation and for a flat space
this requirement is trivially ensured by a proper choice of the
functional form of the analyzing wavelet $\psi(x)$.  However, a
curved space involves an intrinsic parameter having the dimension of
length, namely the local curvature radius.  Therefore, it is
impossible to introduce an exactly self-similar family of wavelets in
a curved space, for example on a sphere.  The intrinsic deviations
from self-similarity are significant when the wavelet scale $a$ is
comparable to or larger than the curvature radius.  Here we suggest a
modification of the wavelet transformation technique to ensure an
approximate self-similarity which becomes almost exact when the scale
of the wavelet $a$ is much smaller than the curvature radius.

Another problem is that one should take special care about condition
(\ref{w-adm}) in numerical realizations of the wavelet
transformation, especially when the data points are irregularly
distributed in space.  It is nontrivial to ensure that the average
value of the wavelet of a scale $a$ is zero when $a$ is comparable to
or smaller than the {\it largest\/} separation of the data points.

The two difficulties mentioned above can be resolved simultaneously
using the {\it gapped\/} (or {\it adaptive\/}) {\it wavelet
technique\/} suggested by Frick et al.\ (1997a) for a one-dimensional
case (see also Frick, Grossmann \& Tchamichian 1998). Here we
generalize this technique to wavelets defined on a spherical surface.

Let us represent the analyzing wavelet in the form
\begin{equation}
\psi(x) = h(x)  \Phi(x)\;,      \label{w_a}
\end{equation}
where $\Phi(x)$ is a positive-definite envelope and $h(x)$ represents
the filling of the wavelet. In the case of the `Mexican hat'
(\ref{hat}) in a plane we have $\Phi(x)=\exp\left(-\half x^2\right)$
and $h(x)=2-x^2$.

Following Frick et al.\ (1997a), we use the wavelet of the form
\begin{equation}
\psi_{a,{\bmath{r}}}({\bmath{r}}') = a^{-1} \left[
h(s/a)-{\Ko}(a,{\bmath{r}}) \right] \Phi(s/a) \label{psi_tild}
\end{equation}
in the numerical realization of the wavelet transformation. Here $s$,
the angular separation of ${\bmath{r}}$ and ${\bmath{r'}}$, is a
function of ${\bmath{r}}=(l,b)$, the position of the centre of the
wavelet and ${\bmath{r}}'=(l',b')$, a current position on the unit
sphere (${\bmath{r}}$ and ${\bmath{r}}'$ are defined on the unit
sphere and should not be confused with radius vectors). The quantity
$\Ko(a,{\bmath{r}})$ is determined from the condition $ \sum_i \psi_i
= 0$, where $\psi_i$ are the values of the wavelet at the data grid
points $(l_i,b_i)$.  This yields
\begin{equation}
\Ko(a,l,b) = \Big[\sum_i \Phi(s_i/a)\Big]^{-1}
    \sum_i h(s_i/a)\Phi(s_i/a)\;,     \label{C}
\end{equation}
where $s_i$ are the distances of the data points to the centre of the
wavelet $(l,b)$. Thus, we have $c=2-\Ko(a,{\bmath{r}})$ in
Eq.~(\ref{hat}).

All the integrals (\ref{wt}) were calculated on the irregular data
grid determined by the particular catalogue using the
Monte-Carlo technique as
\begin{equation}
w(a,l,b) = \Delta S \sum_i \psi_{a,{\bmath{r}}}({\bmath{r}}'_i)
        \,\RM_i \;,
\label{wprim}
\end{equation}
where summation is carried out over the data points. Here $\Delta S=
a^2\left[\sum_j \Phi(s_j/a) \right]^{-1}$ is the area element.

\epsfxsize=8.2cm
\begin{figure}
\centerline{\epsfbox{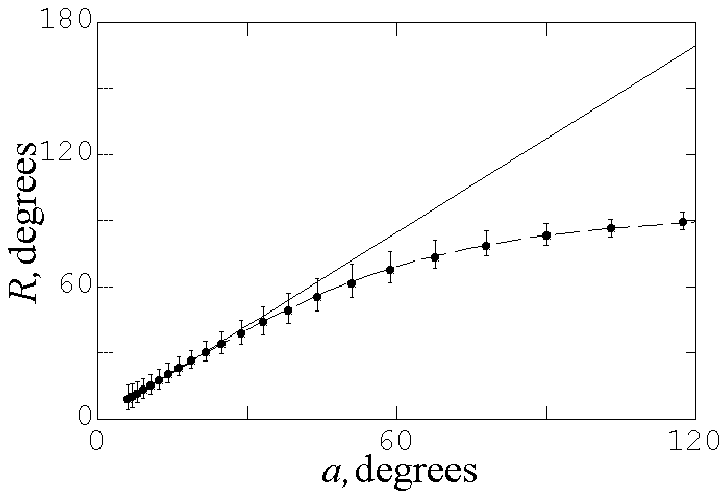}}
\caption[]{
$R$ as a function of
$a$ for a sphere (dashed line) and for a plane (solid).
Error bars indicate discretization errors for the catalogue of
Simard-Normandin et al.\ (1981).}
\label{R_scatter}
\end{figure}

A significant dependence of $\Ko(a,{\bmath{r}})$ on position for
small $a$ affects the scale resolution of the wavelet technique. In
terms of the radius $R$ at which a wavelet has the first zero,
$R=a[2-\Ko(a,{\bmath{r}})]^{1/2}$, the resulting wavelet transform is
unreliable at those scales where the scatter in $R$ arising from the
position variations of $\Ko(a,{\bmath{r}})$ is comparable with $a$.
The specific form of $\Ko(a,{\bmath{r}})$ depends on the distribution
of the data points on the sphere, i.e.\ on the catalogue chosen. In
Fig.~\ref{R_scatter} we show $R$ as a function of $a$ for the
catalogue of Simard-Normandin et al.\ (1981) and indicate the scatter
arising from the inhomogeneous distribution of the data points.  For
$R\la20^\circ$, the relative error exceeds 25\%; we therefore
consider the results obtained from this catalogue to be reliable at
scales $R\ge20^\circ$.  The catalogue of Broten et al.\ (1988)
contains more sources than that of Simard-Normandin et al.\ (1981).
However, the sources in the former catalogue are distributed less
uniformly in the sky, so the resulting discretization error is even
larger than for the latter catalogue.  Therefore, the resolution
limit of both catalogues is adopted as $R=20^\circ$. Note that this
restriction does not apply in Sect.~\ref{LLS} where we consider
low-latitude sources and the separation of data points is reduced by
projection on the equator.

We have tested our implementation of the wavelet transformation using
artificial data sets given on both a regular grid in $(l,b)$ (with a
$81\times41$ mesh) and the grids of the catalogues considered in this
paper. We specified a test distribution of $\RM$ in the sky
consisting of four structures at different scales:  two spots of
different signs having a scale $R=70^\circ$ and located at
$(l,b)=(\pm90^\circ,0)$, one positive spot of a scale $R=40^\circ$ at
$(l,b)=(45^\circ,-45^\circ)$, and another negative spot of
$R=25^\circ$ located at the Galactic Northern pole. The amplitudes of
all the structures were equal to each other. In addition, we
superimposed a white noise of a relative amplitude 15\%.

We calculated the values of $\RM$ of the test distribution at those
positions in the sky where Simard-Normandin et al.\ (1981) give their
values of $\RM$ and also on the grid of the joint catalogue. We then
performed wavelet transformation of the resulting discrete data sets
specified on the irregular grids and checked that the wavelet
transform successfully isolates the structures in the input signal,
and that these structures, if well separated in scale, result in
local maxima of the integral energy spectrum $M(a)$ as given by
Eq.~({\protect\ref{Ma}}) (calculated for the same discrete data set).

This test confirms that our algorithm correctly recovers structures
on real data grids: all the four structures can be clearly seen in
the wavelet transform maps. Traces of the structures at larger
scales can also be seen at smaller scales because our wavelet
provides a good spatial resolution (i.e., it is sensitive to the
position of a structure) and only a moderate scale resolution. The
results remain stable when white noise has been added to the data.

\section*{Acknowledgments}
We are grateful to E.M.~Berkhuijsen for numerous helpful comments and
critical reading of the manuscript, to R.~Beck, J.L.~Han,
P.P.~Kronberg and P.~Reich for useful discussions, to J.P.~Vall\'ee
for providing the updated catalogue of Broten et al.\ (1988), and to
E.~Ryadchenko for his assistance in the compilation of the joint
catalogue.  We thank the anonymous referee for helpful comments.
Financial support from PPARC (grant GR/L30268), the Royal Society
and NATO (grant CRG1530959) is
acknowledged. PF and DS are grateful for the hospitality of the
University of Newcastle where the final version of this paper was
prepared.
AS and DD gratefully acknowledge generous assistance of Miss
A.D.~Sokoloff.


\bsp
\label{lastpage}
\end{document}